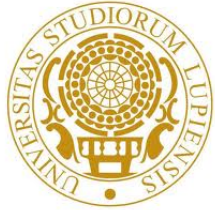

## Università del Salento
Facoltà di Scienze Matematiche, Fisiche e Naturali

Corso di Laurea Magistrale in Fisica

# Relation between Symmetry Groups for Asymptotically Flat Spacetimes

*Relatore:*
Chiar.mo Prof. Luigi Martina

*Laureando:*
Yvonne Calò

Anno Accademico 2016-2017

# Contents

# Aknowledgments

First of all I would like to express my sincere gratitude to my advisor, Prof. Luigi Martina, for the continous support, for his patience and the passion transmitted. His guidance was of fundamental importance during the period of the preparation of the thesis.
I am very grateful to Prof. Peter Horvathy, who gave me the opportunity to approach the issues dealt with in this dissertation, for his hospitality during my Master thesis internship at the Laboratoire de Mathématiques et Physique Théorique of the University of Tours and for the interesting and inspiring discussions.
I thank my previous fellow student Federico Capone for having made time to share what he knows about the BMS group and the related topics and for his critical sensibility.
Finally, I would like to thank also Prof. Marika Taylor and Prof. Claudio Dappiaggi to which I briefly came in contact during this period and with which I had some enlightening discussions.

# Introduction

The concept of *asymptotic flatness* is fundamental in the theory of the General Relativity. In fact, an asymptotically flat spacetime represents in the theory what an isolated system is; then, it allows to define, globally, some physical quantities, such as the total energy and the angular momentum. For this reason, the asymptotic structure of the spacetime has been largely studied over the years.

Particular attention is devoted to the study of the isometries of such spacetimes, namely the spacetime transformations that leave invariant the large distance behavior of the gravitational field or, equivalently, the coordinate transformations that do not change the metric describing the regions far from a massive object. The rigorous mathematical way for describing the isometries of a system is provided by the Group Theory.

It is well known that, for a generic curved spacetime, the Poincaré group plays no role, but we can expect it appears again as the asymptotic symmetry group of asymptotically flat spacetimes. However, it is not exactly what happens. Going to the asymptotic region reached by the massless particles, one finds that the group of isometries is, surprisingly, larger than the Poincaré group. It is the so-called BMS group, developed in the 60s by Bondi, Metzner, van der Burg [1] and Sachs [2] [3].

The BMS group is strictly related to other groups: in particular, in this work of thesis, we are interested in the connection with the Carroll group. Indeed, the latter, and in particular its conformal extension, is demonstrated to be isomorphic to the BMS group [4]. Intuitively, the possible relation between the Carroll group and the BMS group could be deduced from the fact that the first represents the isometry group of a non-causal spacetime, while the other is the symmetry group of the asymptotic region called null infinity and it is also a non-causal region. In the development of the thesis we explain the terminology used here and we show also that the Carroll group is the dual of the Galilei group and depicts a non-relativistic limit of the Poincaré group.

The studying of the properties of the groups related to the BMS group may be useful for the comprehension of the characteristics and the representation theory of the BMS group itself. Such a task is required for several purposes relevant to current research in gravity theory, as we now briefly summarise.

The renewed interest in the BMS group is due to the conjecture that the BMS symmetry rules two completely different physical phenomena, both of them relevant to our understanding of gravity [5]. On the one hand, they provide a fundamental explanation for the so-called observable gravitational memory effect[1], predicted some forty years ago [7] and since then matter of debate in the general relativistic community. On the other hand, the BMS group can be considered the ruler of Weinberg soft theorems for gravitons. Weinberg theorems are general statements about the infrared limit of scattering amplitudes in theories with massless particles, namely long-range interactions. In the context of the investigation first carried on in [8], they arose as Ward identities of (a still to be found) gravitational S-matrix, rooted in the BMS invariance of this operator.

The link of all these topics operated by the concept of asymptotic symmetry group has been stated in an infrared triangular equivalence by Strominger [5]. Such an equivalence can be found also in other gauge theories, thus giving new perspective in understanding their structure. Furthermore, the discussion on the action of the BMS symmetry on the S-matrix seems to be relevant for the implementation of the holographic principle in asymptotically flat spacetimes. The holographic principle, first proposed in [9], states that the degrees of freedom of a gravitational system in a certain volume are encoded in the boundary.

The first and most known (and best posed) example of the holographic principle is considered to be the so-called AdS/CFT correspondence, that is supposed to be an exact duality between a theory of gravity in AdS spacetime and a quantum (conformal) field theory on the boundary.

The flat holography problem consists in finding a holographic description of the physics that may occur in the bulk of flat and asymptotically flat spacetimes. This is a much harder task than the AdS/CFT correspondence as explained in Witten's talk at the conference Strings98 [10] and it may be that we cannot find a similar description to the one available in AdS spacetimes, where two dynamical

---

[1]It will probably be observed with the LISA detector [6], resulting in another striking verification of the predictions of Einstein's theory.

theories -one on the boundary and one in the bulk- are mapped.

Indeed, as we will study in the thesis, the boundary structure of asymptotically flat spacetimes may not allow for a *usual* causal and local quantum field theory to exist. However, the fact that the S-matrix structure is ruled by asymptotic symmetries is, in the way foreseen by Witten, a holographic statement. The sure thing is that the asymptotic symmetry group is bound to play a crucial role in any further consideration around these points. This is also an old lesson from the AdS/CFT correspondence, where the boundary theory is ruled by the same (conformal) group that arises as the symmetry group of the boundary of AdS.

The point is that the properties and the representations (which are crucial for quantum field theory) of the BMS group have not been completely uncovered yet, except for some progress in the case of three-dimensional asymptotically flat spacetimes, where the BMS group can be reconstructed when we take the $\Lambda \to 0$ limit ($\Lambda$ being the cosmological constant) of the Asymptotic Symmetry Group (ASG) of the three-dimensional AdS spacetime [11] [12]. The upshot is that in three dimensions one can study the properties and the representation theory of BMS by referring to the related symmetry group (even though things are not at all straightforward, for example one has to face the problem of non unitary representations).

Guided by these examples one may hope that in order to understand $BMS_4$[2] we can take the indirect way of considering the known properties of the related groups that we have mentioned before. This will be relevant not only for the purposes of developing flat holography but also for understanding gravitational wave memories. In the last chapter we will, indeed, consider the gravitational waves as an example of application, while we do not further discuss neither the fascinating quantum topics nor the issues related to the representations of the BMS group, for brevity reasons.

The plan of the thesis is as follows. In the first chapter we explain the concept of asymptotical flat spacetime and its importance for the theory of the General Relativity. In chapter 2 we introduce the isometry group of an asymptotical flat spacetime at null inifinity, the BMS group, following the two different approaches: the first one reclaims the Penrose geometrical method of the conformal infinity, the second is the first historically found by Bondi *et al.*. In chapter 3 we illustrate the Carroll group both as the contraction of the Poincaré group and as the isometry group of a particular structure. Here we also mention something about the Bargmann structure. In the last chapter we show some applications of the Carroll group. Specifically, we highlight the isomorphism between the conformal Carroll group and the BMS group, the equivalence of the punctured light-cone and the emergence of the Carroll group as the isometry group of the plane gravitational waves. The appendices are just a collection of some useful formulas and concepts.

---

[2]The subscript points out the dimension of the spacetime

# Chapter 1

# Asymptotic flatness

## 1.1 Definition of asymptotically flat spacetime

Unlike other physical theories, such as Newtonian gravitation, in General Relativity (GR) it is not possible to distinguish a non-dynamical background from a field describing the physics of the system. In GR the metric contains both dynamical and non-dynamical informations and can be considered loosely a generalization of the gravitational potential of the Newtonian theory.

The concept of "isolated system", which we are often interested in, is not simple to define and it represents an essential problem to be solved because of its physical implications. Suppose we want to study the structure of a compact object. Obviously it is not isolated from the rest of the universe, but we hope we can ignore the influence of distant matter in order to analyze the massive object as if it was situated in a spacetime which becomes flat far from the compact object. Thus, the fact that our system is not influenced by the distant matter translates into the fact that it can be considered immersed in a flat spacetime. In this sense "isolated system" means "asymptotically flat". Finally, an asymptotically flat spacetime represents the gravitational field near an isolated self-gravitating system.

To give a precise geometrical notion of asymptotic flatness it is necessary to formalize the procedure of taking the limit "as one goes to infinity", or rather to specify the rate at which the metric approaches a Minkowski metric at asymptotic large distances.

Precise fall-off conditions to impose do not exist, but we can follow two guiding

principles:

1. the *fall-off conditions must be not too strong*, because they could be too much restrictive and imply the loss of solutions representing isolated systems;

2. the *fall-off conditions must be not too weak*, because these would allow the presence of too many solutions, which may hidden the useful aspects of asymptotic behavior in a multitude of unphysical solutions.

In view of the above considerations, it is possible to state in formulae how a given metric for a curved spacetime approaches the Minkowski metric at very large distances from the isolated system, namely [15]

$$g_{\mu\nu}(x^\rho) = \eta_{\mu\nu}(x^\rho) + O((x^{\rho_i})^{-m_{\mu\nu}}) \tag{1.1}$$

with the further condition:

$$\partial_{\rho_i} g_{\mu\nu}(x^\rho) = \partial_{\rho_i} \eta_{\mu\nu}(x^\rho) + O((x^{\rho_i})^{-m_{\mu\nu}-1}) \tag{1.2}$$

where $\eta_{\mu\nu}$ is the background metric, in this case the flat metric, and $m_{\mu\nu}$ is a number depending on the dimension of the spacetime and on the asymptotic region investigated, $i^0$ or $\mathscr{I}$ (we will see these regions soon in what follows). Thus, a spacetime is called to be asymptotically flat if there exist a coordinates system such that (1.1) and (1.2) yield.

However, such a definition is not the only one. An alternative definition of asymptotic flatness concerns the addition of a boundary to the spacetime in a suitable way, where the boundary represents the points at infinity. This is an ingenious method to regard infinity and it is the so-called *conformal compactification*. Thus, the points at infinity in the original physical metric turn into finite region in a new metric [16]. This method consists in the creation of a map relating the physical original manifold $(M, g_{\mu\nu})$ with a new manifold $(\tilde{M}, \tilde{g}_{\mu\nu})$ by the formula

$$\tilde{g}_{\mu\nu} = \Omega^2 g_{\mu\nu}, \tag{1.3}$$

with $g_{\mu\nu}$ and $\tilde{g}_{\mu\nu}$ the two metrics and $\Omega$ a smooth function, the so-called conformal factor. It is usual in the literature to denote as *non-physical* the new manifold obtained by the conformal map as seen previously. The map allows to bring the points at infinity to finite regions. The new manifold $\tilde{M}$ represents the original manifold $M$ with the addition of the points at infinity.

We can identify five asymptotic regions for an asymptotically flat spacetime:

- the future null infinity, $\mathscr{I}^+$: set of points which the null rays end at large positive times;

- the past null infinity, $\mathscr{I}^-$: the set of points from which null rays begin;

- future timelike infinity, $i^+$: the set of points approached by timelike geodesics;

- past timelike infinity, $i^-$: asympotic distance from which timelike geodesics come out;

- spacelike infinity, $i^0$: asymptotic structure on a Cauchy surface.

Because of the causality of the space-time it is possible to go to infinity in two different ways: in the null directions and in the timelike directions. Moving in the spacelike direction we do not expect that the spacetime becomes flat at a fixed position at early or late time, since we wish to describe a system representing an isolated body both at early and late times. The $\mathscr{I}^+$ region can be viewed as an idealization of far away observers reached by radiation coming from the isolated system.

The simplest example to show how the conformal compactification works is that of the Minkowski spacetime with metric

$$ds^2 = -dt^2 + dr^2 + r^2 d\omega^2. \tag{1.4}$$

Introducing the retarded time $u = t - r$ and the advanced time $v = t + r$, we set

$$\begin{cases} u = \tan \tilde{U}, & -\frac{\pi}{2} < \tilde{U} < \frac{\pi}{2} \\ v = \tan \tilde{V}, & -\frac{\pi}{2} < \tilde{V} < \frac{\pi}{2} \end{cases}$$

with $\tilde{V} \geqslant \tilde{U}$. Since $r \geqslant 0$, the metric becomes

$$ds^2 = \left(2\cos\tilde{U}\cos\tilde{V}\right)^{-2}\left[-4d\tilde{U}d\tilde{V} + \sin^2\left(\tilde{V} - \tilde{U}\right)d\omega^2\right], \tag{1.5}$$

where $d\omega$ is the infinitesimal solid angle. In these new coordinates the metric approaches the points at infinity for $|\tilde{U}| \to \pi/2$ or $|\tilde{V}| \to \pi/2$. So by choosing

$$\Omega = 2\cos\tilde{U}\cos\tilde{V} \tag{1.6}$$

it is possible to bring the points at infinity to ones in the finite affine parameters space in the new metric

$$d\tilde{s}^2 = \Omega^2 ds^2 = -4d\tilde{U}d\tilde{V} + \sin^2\left(\tilde{V} - \tilde{U}\right)d\omega^2 . \tag{1.7}$$

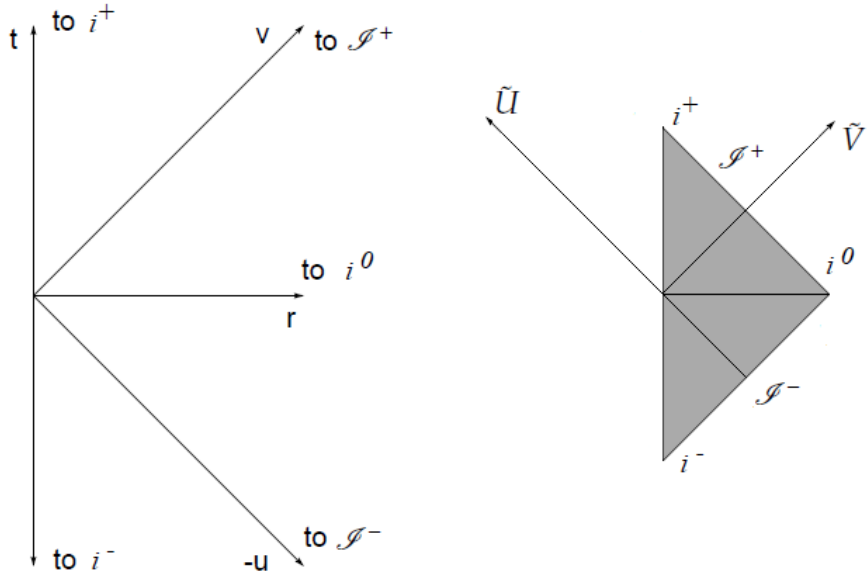

Figure 1.1: On the left, the relation between the $(r,t)$ coordinates and the light-cone coordinates $(u,v)$ and the various asymptotic regions of Minkowski spacetime. On the right the Penrose diagram for Minkowski spacetime.

The relation between $(r,t)$, $(u,v)$ and the various asymptotic regions are (see

figure 1.1):

$$\begin{cases} \mathscr{I}^+ = (|\tilde{U}| \neq \frac{\pi}{2}, \tilde{V} = \frac{\pi}{2}) \Leftrightarrow (u \ finite, v \to \infty) \Leftrightarrow (r \to \infty, t \to \infty, r-t \ finite); \\ \mathscr{I}^- = (\tilde{U} = -\frac{\pi}{2}, |\tilde{V}| \neq \frac{\pi}{2}) \Leftrightarrow (u \to -\infty, v \ finite) \Leftrightarrow (r \to \infty, t \to -\infty, r+t \ finite); \\ i^+ = (\tilde{U} = \frac{\pi}{2}, \tilde{V} = \frac{\pi}{2}) \Leftrightarrow (u \to +\infty, v \to +\infty) \Leftrightarrow (t \to +\infty, r \ finite); \\ i^- = (\tilde{U} = -\frac{\pi}{2}, \tilde{V} = -\frac{\pi}{2}) \Leftrightarrow (u \to -\infty, v \to -\infty) \Leftrightarrow (t \to -\infty, r \ finite); \\ i^0 = (\tilde{U} = -\frac{\pi}{2}, \tilde{V} = \frac{\pi}{2}) \Leftrightarrow (u \to -\infty, v \to \infty) \Leftrightarrow (r \to \infty, t \ finite); \end{cases}$$

In the conformal diagram for Minkowski spacetime, we can identify $\mathscr{I}^+$ with $\mathscr{I}^-$. The definition of conformal infinity of Minkowski spacetime plays an important role in the formulation of the precise asymptotic conditions on physical fields, since they represent the radiation resulting from an isolated source. For more detailed discussion on conformal compactification see for example [17].

It was already mentioned that it is only possible to travel in null and timelike directions. We are interested only to the null infinity.

Now we define formally when a 3 + 1-dimensional spacetime $(M, g_{\mu\nu})$ is asymptotically flat at null infinity. $(M, g_{\mu\nu})$ have to satisfy the following conditions [18]:

1. There exists a function $\Omega$ s.t. $\tilde{g}_{\mu\nu} = \Omega^2 g_{\mu\nu}$ can be smoothly extended to an "unphysical" spacetime $\tilde{M}$ with boundary $\mathscr{I}^+ \cup \mathscr{I}^- \subset \partial \tilde{M}$ and $\mathscr{I}^\pm \cong \Sigma \times \mathbb{R}$, where $\Sigma$ is a compact 2 dimensional manifold;

2. relative to $\tilde{g}_{\mu\nu}$, $\tilde{n} = grad_{\tilde{g}} \Omega$ is a null vector field on $\mathscr{I}^+$, there $d\Omega \neq 0$ holds;

3. the Einstein vacuum equations $Ric_g = 0$ hold for g in an open neighborhood of $\mathscr{I}^+$.

With $\mathscr{I}^\pm \cong \Sigma \times \mathbb{R}$ we do not intend a relation of congruence as usual in mathematics, rather the fact that past and future null infinity have globally the topological structure $\Sigma \times \mathbb{R}$. In fact, emitted from the origin $r = 0$, a massless particle reaches a region which is a sphere at null inifnity and exists such a sphere for every value of the retarded time $u$. The set of all these spheres is precisely $\Sigma \times \mathbb{R}$. Note that the previous conditions defining the asympotic flatness are valid in the case of null infinity only. For an asymptotically flat timelike infinity we have other conditions. For more details about the asymptots see [19] or chapter 11 of [17]. Furthermore, the definition of asymptotic flatness given can be extended in the case of higher even dimensions as discussed in [18].

### 1.1.1 Conformal infinity of the Schwarzschild spacetime

We have seen how to construct the conformal infinity in the case of the flat Minkoskian spacetime. However, for the aims of this thesis, the analysis of the infinity for the Schwarzschild spacetime is more instructive. We start from the standard form of the Schwarzshild metric

$$ds^2 = -\left(1 - \frac{2m}{r}\right)dt^2 + \left(1 - \frac{2m}{r}\right)^{-1} dr^2 + r^2\left(d\theta^2 + \sin^2\theta d\phi^2\right), \tag{1.8}$$

where $m$ is the mass of the spherical simmetric object and $r$ the radial distance. It is evident that the metric (1.8) represents an asymptotically flat spacetime since it reduces to the Minkowski metric if one takes the limit $r \to \infty$, so the asymptotic regions $i_0$ and $\mathscr{I}^\pm$ can be added as in the Minkowskian case.

What is more, for $r = 2m$, known as the Schwarzschild radius, the (1.8) presents a singularity, an hypersurface called the *event horizon*. In the case of a Schwarzschild black hole, the matter and radiation coming from the inner cannot pass the event horizon. Nevertheless, it is not a physical singularity because it depends only on the particular choice of the coordinates.

In order to avoid the problem of the coordinates-dependent singularity and to obtain the past and future null infinity, $\mathscr{I}^-$ and $\mathscr{I}^+$, we introduce the Eddington-Finkelstein coordinates

$$\begin{aligned} u &= t - r - 2m\log(r - 2m) \\ v &= t + r + 2m\log(r - 2m), \end{aligned} \tag{1.9}$$

where $u$ and $v$ are the retarded and advanced time coordinates rispectively. Replacing one at a time the (1.9) in (1.8) we find two new forms for the metric (1.8)

$$\begin{aligned} ds_u^2 &= -\left(1 - \frac{2m}{r}\right)du^2 + 2dudr + r^2\left(d\theta^2 + \sin^2\theta d\phi^2\right) \\ ds_v^2 &= -\left(1 - \frac{2m}{r}\right)dv^2 - 2dvdr + r^2\left(d\theta^2 + \sin^2\theta d\phi^2\right). \end{aligned} \tag{1.10}$$

We are now able to pass from the physical metrics (1.10) to the unphysical ones choosing $\Omega = r^{-1} = w$

$$d\tilde{s}_u^2 = \Omega^2 ds_u^2 = -\left(w^2 - 2mw^3\right)du^2 - 2dudw + d\theta^2 + \sin^2\theta d\phi^2 \quad (1.11)$$
$$d\tilde{s}_v^2 = \Omega^2 ds_v^2 = -\left(w^2 - 2mw^3\right)dv^2 + 2dvdw + d\theta^2 + \sin^2\theta d\phi^2. \quad (1.12)$$

The unphysical metrics (1.11, 1.12) are clearly regular and degenerate for $w = 0$. In particular, the (1.11) allows us to include to the physical spacetime, achieved for $m > 0$, the boundary hypersurface $\mathscr{I}^+$: the future null infinity is the hypersurface obtained from (1.11) for $m = 0$. Analogously, $\mathscr{I}^-$ is given by (1.12) when $m = 0$.

It is important to notice that, in the case of the Schwarzschild spacetime, it is not possible to identify $\mathscr{I}^+$ with $\mathscr{I}^-$, or viceversa. In fact, the extension to negative values of $w$, namely a change from $w$ to $-w$, transforms the metric (1.11) in the (1.12), but with a change in the sign of the mass, $m \to -m$. Thus, the extension across the boundary $\mathscr{I} = \mathscr{I}^+ \cup \mathscr{I}^-$ yields the reversal of the sign of the mass. Since the curvature on $\mathscr{I}$ gives information about the mass, we can state that the curvature on $\mathscr{I}^+$ is opposite to the curvature on $\mathscr{I}^-$. Therefore, it does not make sense to identify the past null infinity with the future null infinity. The conformal diagram is the same as the conformal diagram for Minkowski spacetime, but the points $i^0$ and $i^\pm$ are singular for the conformal geometry. Thus, the conformal infinity is composed of two disjointed hypersurfaces that are topologically two cylinders, $S^2 \times \mathbb{R}$. An artistic way of representing this is as in figure (1.2). The result find for the Schwarzschild represents a model for asymptotically flat spacetimes in general.

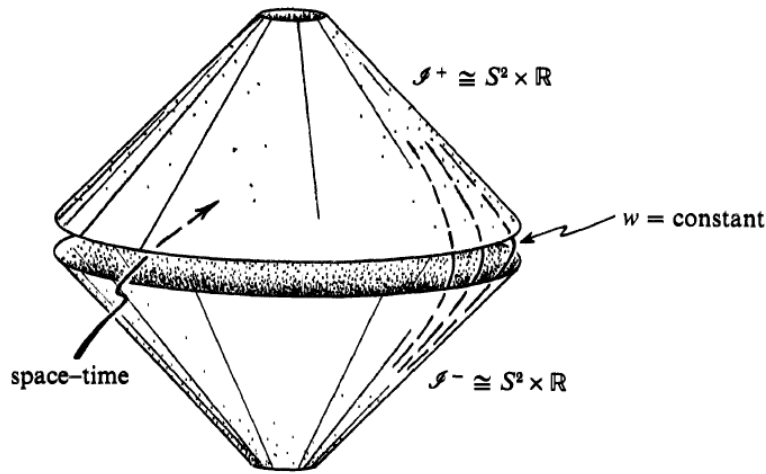

Figure 1.2: Picture of the conformal null infinity for the Schwarzschild spacetime [21].

## 1.2 Asymptotic symmetries in gravitational theory

The group of isometries plays an important role in physics. In particular for Minkowski spacetime this group of isometries is the Poincaré group and it is basic in the standard energy-momentum and angular momentum definitions. Thus, we wish to find a suitable generalization of the concept of isometry group applicable to curved spacetimes approaching Minkowski ones at large distances, i.e. the so-called BMS group; we consider worth of time to clarify some useful concepts.

As it is well known, to each symmetry is associated a conserved quantity as consequence of the Nöther theorem. Let $J^a$ be a vector, such that the continuity equation holds

$$\nabla_a J^a = 0 \tag{1.13}$$

where $\nabla_a$ is the covariant derivative. We can deduce an integral conservation law according to which the integral of the flux of the vector $J^a$ over the boundary $\partial \mathscr{C}$ of a compact region $\mathscr{C}$ must vanish. From Gauss' theorem:

$$\int_{\partial \mathscr{C}} J^a d\sigma_a = \int_{\mathscr{C}} \nabla_a J^a dV = 0, \tag{1.14}$$

where $d\sigma_a$ and $dV$ are respectively the surface and volume elements.

The problem of defining energy and conserved quantities in GR is rather delicate. There are not globally conserved quantities. Beacause of general covariance, the energy-momentum tensor $T^{ab}$ which satisfies the local conservation law

$$\nabla_a T^{ab} = 0. \tag{1.15}$$

This is a non global conservation law because the energy-momentum tensor represents only the matter contribution to the energy ignoring the contribution from the gravitational field energy.

The picture is simplified if the spacetime possesses symmetries, i.e. Killing vectors $\xi^a$. Let $\phi_t : M \to M$ be a one-parameter group of isometries ($\phi_t^* g_{ab} = g_{ab}$);

the necessary and sufficient condition for $\phi_t$ to be a group of isometries is

$$\mathcal{L}_\xi g_{ab} = 0, \tag{1.16}$$

where $\mathcal{L}_\xi$ represents the Lie derivative in direction $\xi$. The necessary and sufficient condition for $\xi$ to be a Killing vector is that its components have to satisfy the system of equations

$$\nabla_a \xi^b + \nabla_b \xi^a = 0. \tag{1.17}$$

Calculation like these may be make faster and clearer by using the formalism of differential forms (see for example [17] ).

Giving vector field $\xi$, it is possible to construct a vector satisfying equation (1.13):

$$P^a = T^{ab} \xi_b; \tag{1.18}$$

in fact, $\nabla_a P^a = \nabla_a T^{ab} \xi_b + T^{ab} \nabla_a \xi_b = 0$: the first term on the right hand side vanishes by (1.15) and the second is zero since $T^{ab}$ is simmetric in the indices. The presence of Killing vectors for the metric implies the existence of integral conservation laws. In the case of the flat Minkowski spacetime there are ten Killing vectors, the generators of the inhomogeneous Lorentz group; four of them generate translations in spacetime

$$L_\alpha = \partial_\alpha \quad , \quad \alpha = 0, 1, 2, 3, \tag{1.19}$$

the other six generate the rotations in spacetime

$$M_{\alpha\beta} = x_\alpha \partial_\beta - x_\beta \partial_\alpha \quad , \quad \alpha = 0, 1, 2, 3. \tag{1.20}$$

It is possible to use those generators to define four vectors $P^a_\alpha$ and six vectors $Q^a_{\alpha\beta}$ which will obey (1.13). We can interpret $P^a_0$ as the flow of energy and $P^a_1$, $P^a_2$, $P^a_3$ as the flow of the three components of the linear momentum.

However, if the metric is not flat it has not Killing vectors in general. Let $q$ be a point on $\mathscr{C}$, one could introduce a set of normal coordinates $x^\alpha$ in a neighbourhood of $q$ such that the components of $g_{ab}$ reduce to those of the flat metric $\eta_{ab}$ and the components of the Christoffel symbols $\Gamma^a_{bc}$ are zero at $q$. However, in a neighbourhood of $q$ $\Gamma^a_{bc}$ differs from zero. Then, $\nabla_{(a} L_{\alpha,b)}$ and $\nabla_{(a} M_{\alpha\beta,b)}$ also differ from zero by an arbitrary small amount. Thus

$$\int_{\partial \mathscr{C}} P_\alpha^b d\sigma_b \quad , \quad \int_{\partial \mathscr{C}} Q_{\alpha\beta}^b d\sigma_b \qquad (1.21)$$

will still be not exactly zero. Integrating over a region whose typical dimensions are very small with respect to the radii of curvature involved in the Riemann tensor $R_{abcd}$, we can get an approximate (local) integral conservation law from (1.15) inspite of an exact conservation law.

Therefore we can think that the curvature of the spacetime gives a non-local contribution to the energy-momentum. This contribution cannot be neglected to obtain a correct integral conservation law.

Finally we can state that exact simmetries do not exist for a generic curved spacetime. In the case of an asymptotic flat spacetime the situation is well defined; "going to infinity" could permit to find Killing vectors useful to define conservation laws. We analize the asymptotic isometry group in the next chapter.

# Chapter 2

# The BMS group

The BMS group is the group of isometries of the asymptotically flat spacetimes at null infinity. It can emerge in two different ways depending on the definition of asymtpotical flat spacetime we use[1].

In this chapter we will study both methods that lead to the BMS group. We will see that the symmetry group of an asymptotically flat spacetime is not the Poincaré group but, surprisingly, a larger group, indeed the BMS group. The Poincaré group is extended by the angle-dependent translations along the null direction, the so-called *supertranslations*.

## 2.1 Conformal Infinity and the BMS group

We start from the study of the BMS group as the group of isometries of a particular geometrical structure imposed on $\mathscr{I}^+$.

From the method of compactification, discussed in the previous chapter, one can derive that a class of equivalent metrics exists on $\mathscr{I}^+$, each of them conformally related to the others. A detailed discussion on the conformal infinity can be found in [21]. We can choose as the representative metric (see for example [21])

$$ds^2_{\mathscr{I}^+} = g_{\mathscr{I}^+} = 0 \times du^2 + d\theta^2 + \sin^2\theta d\phi^2, \tag{2.1}$$

where $u$ is the retarded time. We have deliberately written in (2.1) the piece

---
[1]We have stated in the previous chapter the two definitions of asymptotically flat spacetime, the first one coordinates-dependent and the other coordinates-independent.

$0 \times du^2$ to underline the degeneracy of the metric; when $r \to \infty$, the coefficient of $du^2$ tends to zero. The non-null part of the metric is exactly the standard metric of the sphere.

We can introduce stereographic coordinates $z = e^{i\phi}\cot(\frac{\theta}{2})$, as just seen in the previous chapter, in terms of which the metric on the null infinity becomes

$$g_{\mathscr{I}^+} = 0 \times du^2 + \frac{4}{(1+z\bar{z})^2} dz d\bar{z}. \qquad (2.2)$$

The most general conformal transformation of the compactified plane is

$$z' = \frac{az+b}{cz+d} \qquad (2.3)$$

where a,b,c,d $\in \mathbb{C}$ and can be normalized such that $ad - bc = 1$. The factor in the metric (2.2) transforms as

$$\frac{dz' d\bar{z}'}{(1+zz')^2} = K^2(z,\bar{z}) \frac{dz d\bar{z}}{(1+z\bar{z})^2} \qquad (2.4)$$

and

$$K(z,\bar{z}) = \frac{1+z\bar{z}}{(az+b)(\bar{a}\bar{z}+\bar{b})+(cz+d)(\bar{c}\bar{z}+\bar{d})}. \qquad (2.5)$$

Such transformations (2.3) are the well-known Möbius transformations (also called the fractional linear transformations), and they form a group isomorphic to $PSL(2,\mathbb{C})$, the group of projective special linear transformations of the complex plane.

The Newman-Unti group, which we call NU, is defined as the group of transformations (2.3) and

$$u \to u' = F(u,z,\bar{z}) \qquad (2.6)$$

with $F$ a smooth function on $\mathscr{I}^+$, $\frac{\partial F}{\partial u} > 0$. The transformations given above are the non-reflective motions of $\mathscr{I}^+$ connected to the identity. To be rigorous, we should call this group the restricted NU group, but we will omit this specification.

Consider now the case of the Minkowski spacetime $\mathbb{M}$, which is mapped into itself under the Poincaré transformations. Since the conformal structure of the future null infinity is determined by the conformal structure of $\mathbb{M}$, the metric on $\mathscr{I}^+$ is also invariant under the transformations (2.6) and (2.3). Hence, we can derive that the Poincaré group is a subgroup of the NU. The NU group is clearly much larger than the Poincaré group. In fact the NU is infinite dimensional while, as it is well known, the Poincaré group is ten-dimensional.

The NU group is also much larger than the BMS group. The BMS group is indeed obtained from the NU group with the choice

$$F(u, z, \bar{z}) = uG(z, \bar{z}) + H(z, \bar{z}). \tag{2.7}$$

Such restriction is rigorously found by requiring the preservation of the so-called Bondi parameters, whose definition can be found in [21], [22]. The time $u$ is, in this language, one of the Bondi parameters on each generator of $\mathscr{I}^+$. It is also possible to restrict the form of the function $G$ present in (2.7) by the imposition of an additional structure on $\mathscr{I}^+$ known as *strong conformal geometry*.

There are different ways to introduce the strong conformal geometry for $\mathscr{I}^+$, but we follow that analyzed by Penrose [21].

Let us consider the null tangent vector $\mathcal{N}^a$ to $\mathscr{I}^+$ that is defined starting from $\mathcal{N}_a$ given, near $\mathscr{I}^+$, by

$$\mathcal{N}_a = -\nabla_a \Omega, \tag{2.8}$$

where $\Omega$ is a conformal factor. $\mathcal{N}^a$ transforms under a conformal rescaling with conformal parameter $\Theta$ as follows

$$\mathcal{N}_a \to \mathcal{N}'_a = \Theta \mathcal{N}_a \quad , \quad \mathcal{N}^a \to \mathcal{N}'^a = \Theta^{-1} \mathcal{N}^a \quad on \quad \mathscr{I}^+, \tag{2.9}$$

such that

$$\tilde{g}_{ab} \to \tilde{g}'_{ab} = \Theta^2 \tilde{g}_{ab}, \tag{2.10}$$

where $\tilde{g}_{ab}$ is the *non physical metric* connected to the *physical metric* by $\tilde{g}_{ab} = \Omega^2 g_{ab}$. We have for the line element $dl$ on $\mathscr{I}^+$ the rescaling

$$dl \to dl' = \Theta dl, \tag{2.11}$$

therefore

$$\mathcal{N}^a dl \to \mathcal{N}'^a dl' = \mathcal{N}^a dl \tag{2.12}$$

namely $\mathcal{N}^a dl$ is invariant under the previous scaling transformations. The scaling of the parameter $u$ can be fixed if we choose i.e.

$$\mathcal{N}^a \nabla_a u = 1, \tag{2.13}$$

and with this choice

$$du \to du' = \Theta du. \tag{2.14}$$

All the special parameters $u$ have the same scaling along the generators of $\mathscr{I}^+$. This property of invariance defines what is called the *strong conformal geometry*. Putting together (2.11) and (2.14) we have that

$$\frac{du}{dl}$$

is independent of the choice of $\Theta$. This fact, besides to reflect the property of invariance mentioned before, is useful to define the concept of *null angle*. In order to define the null angle, let us consider two non-null tangent directions at a point $P$ of $\mathscr{I}^+$, specified by $\hat{X}$ and $\hat{Y}$. If no linear combination of $X \in \hat{X}$ and $Y \in \hat{Y}$ is in the tangent null direction at $P$, then the angle between $\hat{X}$ and $\hat{Y}$ is defined by the metric (2.1). However, if the null tangent direction at $P$ is contained in the plane spanned by $\hat{X}$ and $\hat{Y}$, the angle between $\hat{X}$ and $\hat{Y}$ always vanishes. In order to see this, it is possible to introduce a vector $N \in \hat{N}$, where $\hat{N}$ is the null tangent direction at $P$, such that $Y - X = N$. Since $N$ is null, we have

$$0 = g(N, N) = g(Y, Y) + g(X, X) - g(Y, X) - g(X, Y) \tag{2.15}$$

from which we find that the relation determining the angle $\theta$ between the two non-null direction $[X]$ and $[Y]$ using the metric (2.1)

$$\cos\theta = \frac{g(X, Y)}{\sqrt{g(X, X)g(Y, Y)}} = 1 \tag{2.16}$$

and then $\theta$ vanishes.

On the other hand, if we require the strong conformal geometry to hold with the invariance of the ratio $\frac{du}{dl}$, we can define approximately the null angle between two tangent directions at a point of $\mathscr{I}^+$ by

$$\nu = \frac{\delta u}{\delta l}, \tag{2.17}$$

see figure (2.1).

Summarising, the conformal metric (2.2) with the addition of the strong conformal structure on $\mathscr{I}^+$ defines the null angle. Conversely, the conformal metric supplemented by the concept of null angle implies that $\mathscr{I}^+$ is characterized by the strong conformal geometry.

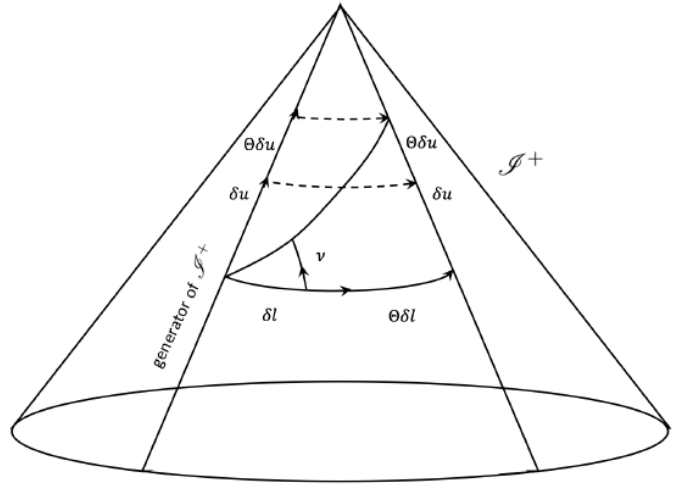

Figure 2.1: A null angle on $\mathscr{I}^+$ is given by $\delta u/\delta l$ and it is defined as the angle between two directions on $\mathscr{I}^+$ whose span contains the null normal direction to $\mathscr{I}^+$. The null angle is invariant under a change of the conformal factor, [23].

Coming back to the transformation (2.14), if we integrate it we obtain for $u$ the following transformation

$$u \to u' = \Theta \left[ u + \alpha(z, \bar{z}) \right] \tag{2.18}$$

and therefore we see that we have obtained the form of the function $F$ by the imposition of the strong conformal geometry. We can identify $G(z, \bar{z})$ with $\Theta(z, \bar{z})$, where

$$G(z, \bar{z}) \equiv \Theta(z, \bar{z}) = \frac{1 + z\bar{z}}{(az + b)(\bar{a}\bar{z} + \bar{b}) + (cz + d)(\bar{c}\bar{z} + \bar{d})} \tag{2.19}$$

in order to preserve the metric under the Möebius transformations. In conclusion we have found that the group preserving the strong conformal geometry, for example preserving both angles [2] and null angles, is the BMS group. A further discussion about the Newman-Unti group and the BMS charge algebra is found in [22].

**Observation.** We have previously introduced, for the two tangent null directions to $\mathscr{I}^+$ in $P$, two angles, $\theta$ and $\nu$. They emerge from the strong conformal geometry imposed on $\mathscr{I}^+$.

But what is their meaning? Well, in order to explain the interpretation of $\theta$ and $\nu$, it is useful to consider directly the geometric structure of the generic spacetime $\mathcal{M}$ rather than the strong conformal geometry of $\mathscr{I}^+$.

We indicate the two non-null tangent directions as $\alpha$ and $\beta$. However, for our purpose it is convenient to start from the orthogonal complements of $\alpha$ and $\beta$, $\alpha^*$ and $\beta^*$. $\alpha^*$ and $\beta^*$ are two timelike hyperplane elements (in fact $\alpha$ and $\beta$ are spacelike).

Looking at the past null cone at $P$, since $\alpha^*$ and $\beta^*$ are timelike, they have non-trivial intersections with the past null cone at $P$ (see figure (2.2)). These intersections form the set of the null directions at $P$. All the null directions at $P$ extend to a ray in $\mathcal{M}$, namely the generator of the past light cone $\mathscr{C}$ of $P$ in $\mathcal{M}$.

By virtue of this observation, we can interpret $\alpha^*$, $\beta^*$, $\alpha$ and $\beta$ in terms of the spacetime geometry of the light rays.

What is more, it is known that a null hypersurface, as the past light-cone $\mathscr{C}$ of $P$, is interpreted spatially as an asymptotically plane wave-front. As a consequence, we have that the 2-dimensional space of the cross-section of $\mathscr{C}$ approaches a Euclidean plane when it moves along $\mathscr{C}$ in future direction.

In the simpler case of Minkowski spacetime $\mathbb{M}$, $\mathscr{C}$ is a null hyperplane and, thus, the 2-space of the cross-section is exactly the Euclidean 2-plane. However, in the general case, we have a 2-space of the cross-section that tends to the Euclidean plane. We indicate this "*limiting* exact Euclidean plane" (following the nomenclature of Penrose [21]) with $\mathbb{E}_p$.

The plane $\mathbb{E}_p$ describes the geometry of the generators of $\mathscr{C}$. The points of the generators of $\mathscr{C}$ are the set of all the null directions at $P$ plus the one generator $\gamma$ of $\mathscr{I}^+$ through $P$, as we can see more clearly from the figure (2.2).

---

[2] the angles on the sphere that are preserved by conformal transformation

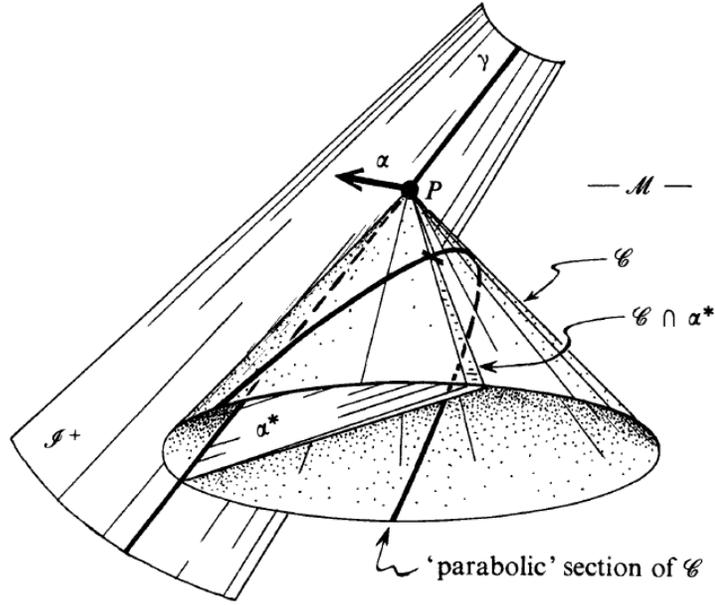

'parabolic' section of $\mathscr{C}$

Figure 2.2: The orthogonal complement $\alpha^*$ of the tangent direction $\alpha$ (to $\mathscr{I}^+$ at $P$) can be used to characterize a parabolic section of the past cone $C$ of $P$, in a straight line in $\mathbb{E}_p$. Conversely, this straight line is used to represent $\alpha$, [21].

The intersection of $\alpha^*$ and $\beta^*$ with $\mathscr{C}$ allows us to pinpoint a parabolic section on $\mathscr{C}$ parallel to $\gamma$. The parabolic section has intrisically an Euclidean 2-metric. Thus, $\mathbb{E}_p$ can be interpreted as this parabolic section in the tangent space $\mathcal{T}[P]$.

We can require the Euclidean 2-metric of the parabolic section to have the correct scaling in agreement with the treatment done. In terms of the unphysical induced metric $\tilde{g}_{ab}$ and the associated covariant vector $N_a$, the equation of the parabolic section in the tangent space $\mathcal{T}[P]$ is

$$x^a N_a + 1 = 0 = \tilde{g}_{ab} x^a x^b \,. \tag{2.20}$$

In this way $\tilde{g}_{ab}$ is invariant under the trasformations (2.9) and (2.10).

When we look at the orthogonal complements of $\alpha^*$ and $\beta^*$, namely $\alpha$ and $\beta$, we have to consider also an element of the 2-plane spanned by the null direction at $P$ and the null direction of $\gamma$, and not just a point of $\mathbb{E}_p$ associated to the null direction at $P$.

Each 2-plane element through the $\gamma$-direction non tangential to $\mathscr{I}^+$ contains this null direction and the null direction through $\gamma$ at $P$. Since the 2-plane element is non tangential to $\mathscr{I}^+$, then its orhogonal complement is tangent to $\mathscr{I}^+$ and does not contain the $\gamma$-direction. As a consequence, the points of $\mathbb{E}_p$ are the non-null tangent 2-plane elements to $\mathscr{I}^+$ at $P$.

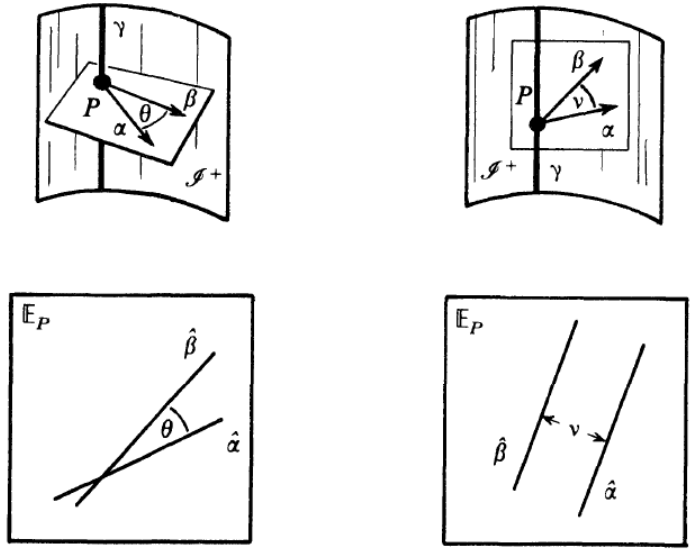

Figure 2.3: Representations of the tangent directions to $\mathscr{I}^+$ in terms of the straight lines in $\mathbb{E}_p$. In the figure on the right it is evident what is the meaning of the null angle $\nu$; in $\mathbb{E}_p$ it corresponds to the Euclidean distance, [21].

At this stage, we can put in evidence the duality between a tangent direction $\alpha$ to $\mathscr{I}^+$ at $P$ with the Euclidean straight line $\hat{\alpha}$ in $\mathbb{E}_p$

$$\mathscr{I}^+ \to \mathbb{E}_p \Rightarrow \alpha \to \hat{\alpha}.$$

Thus, to each non-null tangent direction to $\mathscr{I}^+$ corresponds a straight line in $\mathbb{E}_p$, where the non-null tangent direction to $\mathscr{I}^+$ at $P$ is given by the intersection of two tangent 2-planes to $\mathscr{I}^+$ and a straight line in $\mathbb{E}_p$ is the link of two points in it.

Now, we are able to visualize the two angles $\theta$ and $\nu$ between the two directions $\alpha$ and $\beta$ thanks to the corresponding straight lines $\hat{\alpha}$ and $\hat{\beta}$ in $\mathbb{E}_p$. When the angle $\theta$ between $\alpha$ and $\beta$ is non-zero we find in $\mathbb{E}_p$ that the straight lines $\hat{\alpha}$ and $\hat{\beta}$ intersect among themeselves and the angle they span is $\theta$.

However, it also can happen that $\theta$ is zero. In that case the straight lines $\hat{\alpha}$ and $\hat{\beta}$ in $\mathbb{E}_p$ are parallel. The distance between the parallel straight lines in $\mathbb{E}_p$ is called the Euclidean distance $\nu$. A simple calculation shows that $\nu$ is precisely the null angle between $\alpha$ and $\beta$ in the tangent plane to $\mathscr{I}^+$ at $P$. See figure (2.2).

### 2.1.1 The structure of the BMS group

We have seen how the BMS group emerges as the group of isometries of $\mathscr{I}^+$ equipped with the strong conformal geometry. From this point of view, namely from the geometrical structure of the spacetime, we will underline and deduce some of the properties of the BMS group. We have said several times that the BMS group is much larger than the Poincaré group, but we would understand the meaning of this fact. The BMS group is infinite dimensional and this is due to the presence of the function $H(z,\bar{z})$ in the transformation for $u$ (2.7). When the spacetime is Minkowskian, the function $H$ has the following particular form

$$H(z,\bar{z}) = \frac{H^{00'} - H^{10'}z - H^{01'}\bar{z} + H^{11'}z\bar{z}}{\sqrt{2}(1+z\bar{z})}, \qquad (2.21)$$

where $H^{AB'}$ is a constant and hermitian matrix. This expression holds if $u$ is a special type of Bondi time coordinate in the sense of [21]. The intersection of some light-cone in the Minkowski spacetime $\mathbb{M}$ with the future null infinity gives the coordinate $u = 0$.

The Bondi coordinate $u$ is nothing but the standard retarded time parameter of an inertial obersever in the origin $O$ of the light-cone and time axis $T^a$. This happens if the scaling of the unit celestial sphere associated to $T^a$ is in agreement with the scaling determined by the Bondi time coordinate $u$. Starting from the $u = 0$, all the other values of $u$ are found when the light-cones of the points with all the various position vectors $uT^a$ intersect $\mathscr{I}^+$. See figure (2.4). Therefore, a Bondi time coordinate is a straight line in $\mathbb{M}$ such that $u$ is the proper time along it.

How to find the various transformations characterizing the BMS group? The translations are found when the transformation (2.3) reduces to the identity, thus $G = 1$, and $H(z,\bar{z})$ is of the form given in (2.21). This result is valid non only for $\mathbb{M}$ but also for a generic curved spacetime $\mathcal{M}$.

The set of translations forms a 4-parameter translation subgroup $R$ of the

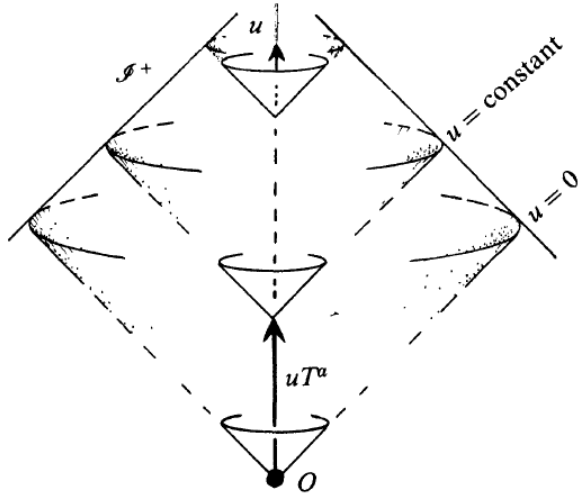

Figure 2.4: Graphic visualization of the Bondi time coordinate, [21]

*BMS* group . On the other hand, if $H(z,\bar{z})$ is a generic function and (2.3) is the identity transformation, we find the so-called supertranslations that form the infinite dimensional subgroup of supertranslations *ST* of the BMS group. Thus *T* is a subgroup both of *BMS* and *ST*. The definitions of translation and supertranslation are independent from the time coordinate *u*. A translation that acts on the supertraslated time Bondi coordinate $u + h(z,\bar{z})$ is the same as a translation that acts on the non transformed *u*. Therefore, the concept of translation is BMS invariant.

The considerations already given can be translated as follows. Let us consider an element **r** of the group of Lorentz rotations *R*, then

$$\mathbf{r}^{-1}T\mathbf{r} = T \tag{2.22}$$

and

$$\mathbf{s}^{-1}T\mathbf{s} = T, \tag{2.23}$$

where $\mathbf{s} \in ST$. *R* is the preserved element since all supertranslations commute.

From the form of the transformation (2.3) we deduce that every element of

*BMS* is of the form **sr**. As consequence of the (2.22) and (2.23) we have

$$\mathbf{b}^{-1}T\mathbf{b} = T \tag{2.24}$$

for every **b** ∈ *BMS*. *T* is a normal subgroup of *BMS*. Similarly, *ST* is a normal subgroup of *BMS*, then

$$\mathbf{b}^{-1}ST\mathbf{b} = ST. \tag{2.25}$$

In conclusion, both translations and supertranslations are BMS invariant.

In order to have a supertranslation, the function *H* must be a generic function of $z$ and $\bar{z}$, then a translation is attained by considering the supertranslation for which *H* is composed only of $l = 0$ and $l = 1$ spherical harmonics (it is always possible to expand *H* in terms of the spherical harmonics). The higher *l*-value part of the *H* expansion is responsible of an important property of the supertranslations: *a pure supertranslation is non Lorentz invariant*. For pure supertranslation we mean "translation-free".

The reason of this property lies on the conformal transformation behaviour of scalars of weight 1. In fact the function $H(z, \bar{z})$ under the action of *R* transforms conformally as a scalar of weight 1. This is because of the conformal transformation property of *u*. The scaling *du* has conformal weight 1 (from $du = \Theta du$) and this is reflected on *H* since *H* transforms conformally in the same way as *u*, as seen from (2.6) and (2.7). For scalars of weight 1 the $l = 0$ and $l = 1$ parts of the expansion transform among themselves under conformal transformations of the 2-sphere, while the higher parts in *l* do not. This is why pure supertranslations are not Lorentz invariant.

The restricted Lorentz transformations are the conformal transformations of the $(z, \bar{z})$-sphere. Thus, the Lorentz group can be seen as a factor group of *BMS*:

$$\mathscr{L} = BMS/ST \tag{2.26}$$

The Lorentz group does not emerge canonically as a subgroup of the BMS group. The subgroup *R* is isomorphic to $\mathscr{L}$, but it is not uniquely identified. In fact, let **s** ∈ *ST*, then

$$R' = \mathbf{s}^{-1}R\mathbf{s} \tag{2.27}$$

is already a subgroup of *BMS* isomorphic to the restricted Lorentz group. However, *R* is the subgroup that leaves invariant a particular cross section of

$\mathscr{I}^+$.

The cross sections of $\mathscr{I}^+$ are also called *cuts*. The cut $\Gamma$ left invariant by the action of $R$ is that defined by $u = 0$, while $R'$ leaves invariant the cut $\Gamma' = \mathbf{s}^{-1}\Gamma$. Only in the case $\mathbf{s} = \mathbf{1}$, $R' \equiv R$ and the invariant cut $\Gamma$ can range over all possible cuts.

We would underline that the restricted Lorentz group arises naturally as a factor group and not as a subgroup of the BMS group, while the Poincaré group arises naturally as a subgroup of *BMS*. In order to see that, consider the simple case of the Minkowski spacetime $\mathbb{M}$. Thus, the Poincaré $P$ is a subgroup of *BMS* whose generators are translations $T$ and rotations $R$.

Again the Lorentz rotation groups are not singled out and it is the subgroup of *BMS* leaving invariant certain cuts. The cuts left invariant by $R$ are the so-called good cuts and they are the intersection of the light-cone, which originates in some event on $\mathbb{M}$, with $\mathscr{I}^+$. Any good cut is related to the others by a translation. The effect of a supertranslation on a good cut is to transform it into a bad cut. Finally, a good cut is not BMS invariant.

In conclusion, we can state that many copies of the Poincaré group exist. In fact if $\mathbf{s}$ is not an element of $R$, then $R'$ is not an element of $P$ and a new 10-parameter subgroup of *BMS* emerges:

$$P' = \mathbf{s}^{-1}P\mathbf{s}. \tag{2.28}$$

$P'$ is isomorphic to $P$, but it is not the same group. $P'$ and $P$ are joined by the subgroup $R$ of supertranslations.

## 2.2 The Bondi-Metzner-Sachs method

In the second section of this chapter we study the BMS group as originally found by Bondi, Metzner, van der Burg and Sachs in [1] and [2]. In the development of this topic we follow the planning of [24].

Let us start from the Minkowski flat metric:

$$ds^2 = \eta_{ab}dx^a dx^b = -dt^2 + dx^2 + dy^2 + dz^2; \tag{2.29}$$

it is convenient to rewrite the metric in terms of the retarded time $u$ and the spherical coordinates $r, \theta, \phi$ defined as follows:

$$u = t - r, \quad r\cos\theta = z, \quad r\sin\theta e^{i\phi} = x + iy, \tag{2.30}$$

2.29 becomes:
$$ds^2 = -du^2 - 2dudr + r^2(d\theta^2 + \sin^2\theta d\phi^2). \tag{2.31}$$

Introducing the metric on the unit $2-sphere$ $\gamma_{AB} = diag(1, \sin^2\theta)$, (2.31) is:

$$ds^2 = -du^2 - 2dudr + r^2\gamma_{AB}dx^A dx^B, \tag{2.32}$$

with $A, B$ labelling the angle variables $\theta$ and $\phi$. The coordinates (2.30) are characterized as follows:

- $u = const$ defines null hypersurfaces, everywhere tangent to the light-cone, since the associated normal co-vector $k_a = \nabla_a u$ is null;

- $r$ is the luminosity distance and it is defined so that the area of the surface element given by $u = const, r = const$, is $r^2 \sin\theta d\theta d\phi$;

- a *ray* is the line with tangent $k^a = g^{ab}\nabla_b u$ and along which $\theta$ and $\phi$ are constant.

Now we want to generalize the case of flat Minkowski spacetime to that of a generic spacetime, requiring to have similar properties to those just listed. Sachs [2], generalizing the work of Bondi, Metzner and van der Burg [1], asserts that, for an asymptotically flat spacetime, at least one set of coordinates, ($u = x^0, r = x^1, \theta = x^2, \phi = x^3$), exists with similar behaviour to the previous coordinates used for the Minkowski spacetime. $u$ is no longer the retarded time $t - r$, but a properly retarded time such that $u = const$ defines null hypersurfaces. Defining the normal vector $k^a = g^{ab}\nabla_b u$, with $k^a k_a = 0$, and using $g_{ab}g^{bc} = \delta_a^c$, we have $g^{uu} = 0$. The coordinate $r$ has to measure the distance of the null hypersurfaces from the origin satisfying $det(g_{AB}) = r^4 det(h_{AB}) = r^4 b(u, x^1, x^2)$. As in the previous case $A, B, ... = 2, 3$ specify the angle variables. Since $x^A$ are constant on the integral curves of $k^a$, namely $k^a \nabla_a x^A = 0$, then $g^{uA} = 0$. From the fact that the only coordinate which varies along the ray is $r$, Bondi gives the restriction $g_{rr} = 0$. We can easily verify the condition $g_{rA} = 0$. In fact $\delta_A^u = g^{uC}g_{CA} = g^{uB}g_{BA} + g^{ur}g_{rA}$, but we have just shown that $g^{uB}, g_{BA}, g^{ur} \neq 0$, from which $g_{rA} = 0$. For this set of coordinates, the line element around each point of the manifold takes the form:

$$ds^2 = e^{2\beta}\frac{V}{r}du^2 - 2e^{2\beta}dudr + g_{AB}(dx^A - U^A du)(dx^B - U^B du), \quad (2.33)$$

where $g_{AB} = r^2 h_{AB}$ and

$$2h_{AB}dx^A dx^B = (e^{2\gamma} + e^{2\delta})d\theta^2 + 4\sin\theta \sinh(\gamma - \delta)d\theta d\phi$$
$$+ (\sin\theta)^2(e^{-2\gamma} + e^{-2\delta})d\phi^2, \quad (2.34)$$

so it turns out $det(h_{AB}) = (\sin\theta)^2$. $V, U^A, \beta, \gamma, \delta$ are six functions depending on the coordinates. We require that the metric (2.33) reduces to (2.31) for large $r$:

$$\lim_{r\to\infty}(ds^2) = -du^2 - 2dudr + r^2\gamma_{AB}dx^A dx^B = ds_m^2. \quad (2.35)$$

In order to ensure (2.35), three conditions are sufficient [3]:

1. all the metric components and the other quantities involved can be expanded in powers of $\frac{1}{r}$ with at most one finite pole at $r = \infty$ in the ranges $u_0 \le u \le u_1$, $r_0 \le r \le \infty$, $0 \le \theta \le \pi$ and $0 \le \phi \le 2\pi$.

2. it is possible to take the limit $r \to \infty$ for some choices of $u$ in the range $u_0 \le u \le u_1$;

3. for the same choice of $u$ and for some choice of $\theta$ and $\phi$ we require

$$\lim_{r\to\infty}(\frac{V}{r}) = -1, \quad \lim_{r\to\infty}(rU^A) = \lim_{r\to\infty}\beta = \lim_{r\to\infty}\gamma = \lim_{r\to\infty}\delta = 0, \quad (2.36)$$

Thus we can write the metric for large $r$ as $ds^2 = ds_M^2 + corrections$ with the appropriate fall-off conditions [25]

$$g_{AB} = r^2 h_{AB} = r^2(\gamma_{AB} + O(r^{-1})), \quad \beta = O(r^{-2}), \quad U^A = O(r^{-2}), \quad (2.37)$$

and hence

$$g_{AB} = r^2 h_{AB} + O(r) \quad g_{uu} = -1 + O(r^{-1}) \quad g_{ur} = -1 + O(r^{-2}) \quad g_{uA} = O(1). \quad (2.38)$$

Given the previous asymptotic behaviour (2.37), it is not trivial to find the precise coefficients of the expansion of the functions present in the metric (2.33).

In the next paragraph we clarify what those coefficients are and how to find their expressions.

### 2.2.1 The boundary conditions

Referring to the principles (1.1, 1.2) we can assume the following expansion for the metric:

$$g_{AB} = r^2 h_{AB} = r^2(\gamma_{AB} + \frac{1}{r}C_{AB} + \frac{1}{r^2}D_{AB} + \frac{1}{r^3}E_{AB} + O(r^{-4})), \tag{2.39}$$

in agreement with the asymptotic behavior of gravitational waves. $g_{AB}$ is the part of the metric spanned by the two coordinates $(x^1, x^2)$ on $\mathscr{I}$ and $\gamma_{AB}$ the metric of the unit 2–sphere.

It is possible to determine the expansion of the functions $V, U^A, \beta, \gamma, \delta$ on which the metric (2.33) depends. We start from the Einstein vacuum field equations $G_{ab} = R_{ab} - \frac{1}{2}Rg_{ab} = 0$ written down in Bondi coordinates. The process of solving the Einstein equations for asymptotically flat spacetimes is shown in more details in [1]. According to the Bianchi identy, we find the following equations, called the *main equations*, for the various components of the Ricci tensor $R_{ab}$:

$$\begin{cases} (I) \ R_{rr} = 0 \\ (II) \ R_{rA} = 0 \\ (III) \ R_{AB} = 0 \Rightarrow \begin{cases} g^{AB}R_{AB} = 0 & \text{trace part} \\ g^{CA}R_{AB} = 0 & \text{traceless part.} \end{cases} \end{cases}$$

Solving the main equations implies automatically that other two *supplementary equations* hold:

$$\begin{cases} (a) \ R_{uu} = 0 \\ (b) \ R_{Au} = 0 \end{cases}$$

with the *trivial equation*

$$R_{ur} = 0.$$

The (I) of the main equations is a differential equation for $\beta$ and the (II) a differential equation for $U^A$. In the iterative procedure to solve the main equations, $C_{AB}$ is assumed as free data. The main equations allow to establish the first non-vanishing order of the expansion of the functions $\beta$, $U^A$ and $V$. As we can see in [1], the integration of the main equations determine the functions present in the Bondi gauge apart from some integral functions.

At the lowest order, the integration with respect to $r$ of $R_{rA}$ gives, as an integration constant, the so-called angular momentum aspect $N^A = N^A(u, x^C)$. The integration with respect to $r$ of the trace part of (III), $g^{AB}R_{AB} = 0$, gives, as an integration constant, the *Bondi mass aspect*, $m = m(u, x^C)$. The integration constant from $R_{rr} = 0$ is set to zero.

The expansions are:

$$\beta = -\frac{1}{r^2}\left(\frac{1}{32}C_{AB}C^{AB}\right) - \frac{1}{r^3}\left(\frac{1}{12}C_{AB}{}^{AB}\right) + O(r^{-3}); \tag{2.40}$$

$$\frac{V}{r} = -1 + \frac{2m}{r} + \frac{2M}{r^2} + O(r^{-3}); \tag{2.41}$$

$$U^A = \frac{S^A}{r^2} + \frac{1}{r^3}\left[-\frac{2}{3}N^A + \frac{1}{16}D^A(C_{BC}C^{BC}) + \frac{1}{2}C^{AB}D^C C_{BC}\right] + O(r^{-4}). \tag{2.42}$$

We denote with $D^A$ the covariant derivative associated with $\gamma_{AB}$. Imposing the gauge condition $\partial_r \det(h_{AB}) = 0$, we have the constraints:

$$\gamma^{AB}C_{AB} = 0, \tag{2.43}$$

$$D_{AB} = C_{CD}C^{CD}\frac{\gamma_{AB}}{4} + \mathscr{D}_{AB}, \tag{2.44}$$

$$E_{AB} = \frac{\gamma_{AB}}{2} + \mathscr{E}_{AB}, \tag{2.45}$$

where $\mathscr{D}_{AB}$ and $\mathscr{E}_{AB}$ are two traceless tensors. The traceless part $g^{CA}R_{AB} = 0$ is useful to determine the $u$-evolution of the subleading terms in the expansion of the metric in terms of the known functions $\beta$, $U^A$ and $V$. As just mentioned, $C_{AB}$ is fixed as an initial data since its $u$-evolution is not given by $g^{CA}R_{AB} = 0$.

The supplementary equations determine the $u$-evolution of the Bondi mass and angular momentum aspect, respectively

$$\partial_u m = \frac{1}{4} D_A D_B N^{AB} - \frac{1}{8} N_{AB} N^{AB}, \tag{2.46}$$

$$\partial_u N_A = D_A m + \frac{1}{4}(D_B D_A D_C C^{BC} - D^2 D^C C_{CA}) + \frac{1}{4} D_B (N^{BC} C_{CA} +$$

$$+ 2 D_B N^{BC} C_{CA}), \tag{2.47}$$

where $N_{AB} := \partial_u C_{AB}$. In the case of non-vacum Einstein equations $G_{\mu\nu} := R_{\mu\nu} - \frac{1}{2} R g_{\mu\nu} = 8\pi T_{\mu\nu}$ the same procedure is used. In addition to what has been done up until now, it is necessary to impose suitable asymptotic conditions on the stress-energy tensor $T_{\mu\nu}$. For more details see [26], [23].

### 2.2.2 Isometries of the BMS group

Now we have sufficient elements to characterize the isometries of an asymptotically flat spacetime. It is precisely the set of those diffeomorphisms of future null infinity, $\mathscr{I}^+$ that is called BMS group.

The isometries of a manifold are generated by the Killing vector fields, as just mentioned before. So, in this case, we look for asymptotic Killing vector fields preserving the Bondi gauge conditions ($g_{rr} = g_{rA} = 0$, $det(g_{AB}) = r det(h_{AB})$) and the leading terms in the metric expansion (2.38).

From differential geometry it is well known that the infinitesimal change of the metric tensor is given by the Lie derivative in the direction of the Killing vector $\xi_a$:

$$\delta g_{ab} = \mathcal{L}_\xi g_{ab}. \tag{2.48}$$

In our case, the variations of the components of the metric are given by

$$\mathcal{L}_\xi g_{rr} = 0, \quad \mathcal{L}_\xi g_{rA} = 0, \quad g^{AB} \mathcal{L}_\xi g_{AB} = 0, \tag{2.49}$$

$$\mathcal{L}_\xi g_{uu} = O(r^{-1}), \quad \mathcal{L}_\xi g_{ur} = O(r^{-2}), \quad \mathcal{L}_\xi g_{uA} = O(1), \quad \mathcal{L}_\xi g_{AB} = O(r). \tag{2.50}$$

The equations (2.49) are the gauge-preserving conditions and are obtained putting together the Bondi gauge conditions and the (2.48). The (2.50) are the boundary-preserving conditions and come from (2.38).

From (2.49) and (2.50) it is not too difficult to find the Killing vector field.

Let us consider the first of (2.49)

$$\mathcal{L}_\xi g_{rr} = \xi^\mu \partial_\mu g_{rr} + (\partial_r \xi^\mu) g_{\mu r} + (\partial_r \xi^\mu) g_{r\mu} = 0; \qquad (2.51)$$

since $g_{rr}$ is zero and $g_{\mu r}$ is different from zero only for $\mu = u$, then

$$\mathcal{L}_\xi g_{rr} = 2e^{2\beta} \partial_r \xi^u = 0 \quad \Rightarrow \quad \xi^u = f(u, x^A), \qquad (2.52)$$

with $f(u, x^A)$ an integration function with respect to $r$. From the second of the (2.49) we obtain the expression for $\xi^A$, in fact

$$\mathcal{L}_\xi g_{rA} = \xi^\mu \partial_\mu g_{rA} + (\partial_r \xi^\mu) g_{\mu A} + (\partial_A \xi^\mu) g_{r\mu} = 0 \qquad (2.53)$$

and since $g_{rA} = 0$, then

$$\partial_r \xi^B g_{BA} + \partial_A \xi^u g_{ru} = 0$$
$$\Rightarrow \partial_r \xi^B g_{BA} - e^{2\beta} \partial_A f = 0$$
$$\Rightarrow \xi^A = Y^A(u, x^B) - \partial_B f \int_r^\infty dR e^{2\beta} g^{AB} \qquad (2.54)$$

with $Y^A$ an integration function with respect to $r$. The expression of $\xi^r$ can be found starting from the last of (2.49) that is an algebraic equations for $\xi^r$. So we have

$$\xi^r = -\frac{r}{2}(D^A \xi_A - U^C \partial_C f), \qquad (2.55)$$

where $D^A$ is again the covariant derivative associated to $h_{AB}$. The full asymptotic Killing vector field is obtained combining (2.54), (2.55) and (2.52), thus

$$\vec{\xi} = f \partial_u - \frac{r}{2}(D_A \xi^A - U^C \partial_C f) \partial_r + [Y^A(u, x^B) - \partial_B f \int_r^\infty dR e^{2\beta} g^{AB}] \partial_A. \qquad (2.56)$$

The equations (2.50) written in explicit form carry three relations for the functions $f(u, x^A)$ and $Y^A$ from which it is possible to deduce the expression of the functions themselves

$$\partial_u f = D_A \frac{Y^A}{2}, \quad \partial_u Y^A = 0 \qquad (2.57)$$

$$D_A Y_B + D_B Y_A - D_C Y^C h_{AB} = 0 \quad \Rightarrow \quad 2D_{(A} Y_{B)} = \psi h_{AB}, \qquad (2.58)$$

where $\psi := D_C Y^C$. From the second of the (2.57), we deduce that $Y^A$ is independent from $u$ and hence it is a vector on the 2-sphere only. Specifically, it is a conformal Killing vector on $S^2$ which generates a conformal transformation.

The procedure yielding (2.57) from (2.50) is described in detail in [27], [23]. We mention here only the fundamental steps. We start computing $\mathcal{L}_\xi g_{AB}$. There is a relation between the Lie derivative and the Christoffel symbols, $\mathcal{L}_\xi g_{AB} = \nabla_A \xi_B + \nabla_B \xi_A = \partial_A \xi_B + \partial_B \xi_A - 2\Gamma^u_{AB} - 2\Gamma^r_{AB}\xi_r - 2\Gamma^C_{AB}\xi_C$. Now, using the expansions (2.38), it is possible to find the expansion of the Christoffel symbols and consequently an expanded form of the Lie derivative.

Thus, we obtain an expansion in powers of $r$. At this stage we have to set to zero all the terms with a wrong power of $r$. According to equations (2.50) we know that $\mathcal{L}_\xi g_{AB} = O(r)$, then we have to set to zero all the terms with a power of $r \geqslant 2$. After calculations one finds that the only non-trivially vanishing term is the term of the second order in $r$.

After this short digression we return to (2.57). Integrating the first of (2.57) we obtain an expression for the function $f(u, x^A)$,

$$f(u, x^A) = \alpha(x^A) + uD_A \frac{Y^A}{2}, \qquad (2.59)$$

with $\alpha(x^A)$ an arbitrary function of the angles that can be parametrized by

$$\alpha = t^\mu n_\mu + \sum_{l=2}^{\infty} \sum_m \alpha^l_m \mathcal{Y}^l_m \qquad (2.60)$$

where $t^\mu n_\mu = (t^0, t^i)$, $n^\mu = (1, n^i)$ and $\mathcal{Y}^l_m$ are spherical armonics.

Equation (2.56) with (2.59) represents the most general transformation that leaves invariant the Bondi metric and the fall-off conditions. This transformation depends on three functions of the angles, $f$ and $Y^A$, and on the metric components. Expanding it to first order in $1/r$

$$\vec{\xi} = \left[\alpha(x^A) + uD_A \frac{Y^A(x^B)}{2}\right]\partial_u + Y^A \partial_A - \frac{1}{2}(rD_A Y^A - D^2 f)\partial_r + O(r^{-1}); \qquad (2.61)$$

it has the right scaling with $r$ and is valid in all the spacetime. The expression

(2.61) reduces at $\mathscr{I}$ as

$$\vec{\xi} = \xi^u \partial_u + \xi^A \partial_A = \left[\alpha(x^A) + uD_A \frac{Y^A(x^B)}{2}\right]\partial_u + Y^A \partial_A. \tag{2.62}$$

The last equation represents the infinitesimal form of the transformations characterizing the BMS group.

Now, we write the finite form of the transformations that leave invariant the asymptotic form of the Bondi metric. For the set of angular coordinates $x^a$ we have that $\theta$ and $\phi$ transform under finite conformal transformations as

$$\theta \to \bar{\theta} = H(\theta, \phi)$$
$$\phi \to \bar{\phi} = I(\theta, \phi), \tag{2.63}$$

with

$$d\bar{\theta} + sin^2\bar{\theta}d\bar{\phi}^2 = K^2(\theta, \phi)(d\theta^2 + sin^2\theta d\phi^2) \tag{2.64}$$

and hence

$$K^4 = J^2(\theta, \phi; \bar{\theta}, \bar{\phi})sin^2\theta(sin\bar{\theta})^{-2}, \tag{2.65}$$

where $J$ is the Jacobian

$$J = det\begin{pmatrix} \frac{\partial H}{\partial \theta} & \frac{\partial H}{\partial \phi} \\ \frac{\partial I}{\partial \theta} & \frac{\partial I}{\partial \phi} \end{pmatrix}$$

and $H$, $I$ and $K$ are conformal functions. The finite form of the transformation for $u$ is

$$u \to \bar{u} = K^{-1}[u + \alpha(\theta, \phi)]. \tag{2.66}$$

Note in (2.62) that $\alpha(x^A)$ and $Y^A(x^B)$ do not mix so we can write the generators of the isometries, the Killing vectors, as the sum of two pieces $\vec{\xi} = \vec{\xi}_\alpha + \vec{\xi}_Y$, where

$$\vec{\xi}_\alpha = \alpha(x^A)\partial_u \tag{2.67}$$
$$\vec{\xi}_Y = \left[uD_A \frac{Y^A(x^B)}{2}\right]\partial_u + Y^A \partial_A. \tag{2.68}$$

The (2.67) is the generator of the supertranslations and the (2.68) is the generator of the conformal transformations on the 2-sphere. These transformations constitute the BMS group.

From the last consideration we deduce the aforementioned fact that the simmetry group of an asymptotically flat spacetime is not the Poicaré group but the larger BMS group which is composed by the abelian infinite dimensional subgroup of supertranslations and a finite dimensional non-abelian group of the conformal transformations. In particular, from (2.67) and (2.68) we can state that the BMS algebra is the semi-sum of those two groups and hence the BMS group is their semi-direct product.

Various extensions of the BMS group exist. In order to give a brief description of these extensions of the BMS group it is convenient to introduce the Bondi gauge conditions in stereographic coordinates. We see this in the next section.

### 2.2.3 Bondi conditions in stereographic coordinates

It is often useful to pass to complex coordinates. In particular, in our case, this change of coordinates helps to understand in a more evident way the action of symmetries on $S^2$ at null infinity.

The passage is from $(\theta, \phi)$ to $(z, \bar{z})$ through the following transformations:

$$\begin{cases} z = e^{i\phi} cot(\frac{\theta}{2}) \\ \bar{z} = e^{-i\phi} cot(\frac{\theta}{2}). \end{cases}$$

First of all, we calculate from these transformations $dz$ and $d\bar{z}$ in order to find the line element relative to the metric of the unit 2-sphere $d\Lambda^2 = (d\theta^2) + sin^2\theta(d\phi^2)$. We have

$$dz = \frac{ie^{i\phi} \sin\theta d\phi - e^{i\phi} d\theta}{2sin^2\frac{\theta}{2}}, \quad d\bar{z} = -\frac{ie^{-i\phi} \sin\theta d\phi - e^{-i\phi} d\theta}{2sin^2\frac{\theta}{2}}, \quad (2.69)$$

from which $dzd\bar{z} = \frac{d\Lambda^2}{4} sin^{-2}(\frac{\theta}{2})$. It is easily verified $dzd\bar{z} = \frac{d\Lambda^2}{4} sin^{-2}$ and since $z\bar{z} = cot^2(\frac{\theta}{2}) \Leftrightarrow 1 + z\bar{z} = \frac{1}{sin^2\frac{\theta}{2}}$ we can write:

$$d\Lambda^2 = h_{AB}dx^A dx^B = \frac{2dzd\bar{z}}{(1+z\bar{z})^2} + \frac{2dzd\bar{z}}{(1+z\bar{z})^2} := 2\gamma_{z\bar{z}}dzd\bar{z}, \quad (2.70)$$

where $\gamma_{z\bar{z}} = \frac{2}{(1+z\bar{z})^2}$ which is the metric of the unit sphere in stereographic coordinates. There are two poles: $z = 0, z = \infty$.

The metric in the new coordinates $(u, r, z, \bar{z})$ is [28], [29]

$$ds^2 = -du^2 - 2dudr + 2r^2\gamma_{z\bar{z}}dzd\bar{z} + \frac{2m_B}{r}du^2 + rC_{zz}dz^2 + c.c. + D^zC_{zz}dudz + c.c. + O(r^{-2}). \tag{2.71}$$

The functions $U_z$ and $U_{\bar{z}}$ at the lowest order have the following constraints

$$U_z = -\frac{1}{2}D^z C_{zz} \quad U_{\bar{z}} = -\frac{1}{2}D^{\bar{z}}C_{\bar{z}\bar{z}}, \tag{2.72}$$

where $D^z$ is the covariant derivative with respect to the metric on the 2-sphere in complex variables, the tensor $C_{AB}$, symmetric and traceless, has components $C_{zz}$ and $C_{\bar{z}\bar{z}}$.

The infinitesimal transformations that leave the metric invariant are

$$u \to u - f, \quad r \to r - D^z D_z f$$
$$z \to z + \frac{1}{r}D^z f, \quad \bar{z} \to \bar{z} + \frac{1}{r}D^{\bar{z}}f, \quad f = f(z, \bar{z}).$$

The Killing vector is

$$\xi_f = f\partial_u + D^z D_z f \partial_r - \frac{1}{r}\left(D^{\bar{z}}f\partial_z + D^z f\partial_{\bar{z}}\right). \tag{2.73}$$

We can also write the expression for the *Bondi mass aspect* $m_B$ and the Bondi *news tensor*:

$$\partial_u m = \frac{1}{4}\partial_u\left(D_z^2 C^{zz} + D_{\bar{z}}^2 C^{\bar{z}\bar{z}}\right) - T_{uu}$$
$$\partial_u N_z = \frac{1}{4}\partial_z\left[D_z^2 C^{zz} - D_{\bar{z}}^2 C^{\bar{z}\bar{z}}\right] + \partial_z m + R_{uz};$$

where $T_{uu} = \frac{1}{4}N_{zz}N^{zz} + 4\pi \lim_{r\to\infty}(r^2 T_{uu}^M)$ is the total energy flux through a given point on $\mathscr{I}$ and $N_{zz} = \partial_u C_{zz}$.

## 2.3 The Global BMS group

We have already justified the structure of the BMS group as the semi-direct product of two subgroups. The global BMS group is the semi-direct product of the infinite abelian subgroup of supertranslations (ST) and a non-abelian group isomorphic to the orthochronous Lorentz group $L^\uparrow = SO(1,3)$

$$BMS^{glob} = ST \ltimes L^\uparrow. \tag{2.74}$$

The BMS group with this structure is how it was originally discovered. It is found by requiring for the function $\alpha(x^A)$ and the conformal Killing vector field on $S^2$, $Y^A$, to admit an expansion in terms of spherical harmonics and vector spherical harmonics respectively. Hence, the transformations defining the BMS group are globally well defined. By virtue of this expansion in spherical harmonics we can write, following Nichols and Flanagan [26], the conformal Killing vector field solution of (2.57), as the sum of two pieces

$$Y^A = D^A \chi + \epsilon^{AB} D_B \kappa =: E^A + S^A, \tag{2.75}$$

where $\chi$ and $\kappa$ are spherical harmonics with $l = 1$. Thus, the six solutions for $Y^A$ can be interpreted as the sum of three solutions with electric parity and three solutions with magnetic parity. The former correspond to translations and the latter to rotations. The distinction in electric parity piece and magnetic parity piece is not so important in this dissertation. We have just mentioned that they are important for the charges associated to the BMS algebra. The charges associated to the electric parity piece, super center-of-mass charges, and those associated to the magnetic parity piece, superspin charges, have very different physical interpretations and characteristics.

Let's neglect the charges for the moment. We would like to find the relation between the conformal group of the 2-sphere at the null infinity with the Lorentz group. It is convenient to work in stereographic coordinates. In fact it is well known from the complex analysis that the Möbius group, the group of conformal transformations of the celestial sphere, is isomorphic to the group $SO(1,3)$.

The conformal Killing equations in stereographic coordinates are

$$\partial_z Y^{\bar{z}} = 0 \quad \partial_{\bar{z}} Y^z = 0. \tag{2.76}$$

These equations express the fact that

$$Y^{\bar{z}} := Y^{\bar{z}}(\bar{z}) \quad Y^z := Y^z(z). \tag{2.77}$$

Bondi and Sachs assumed that this function was analytical rather than arbitrary meromorphic function of its argument. The analyticity is guaranteed if one assumes that they have an expansion in terms of spherical harmonics of the following form

$$Y^z(z) = \frac{az+b}{cz+d} \quad with \quad det\begin{pmatrix} a & b \\ c & d \end{pmatrix} \neq 0, \quad a,b,c,d \in \mathbb{C}. \tag{2.78}$$

These transformations have a zero for $z = -b/c$ and a pole for $z = -d/c$. The same remains valid for $Y^{\bar{z}}$. The (2.78) is the set of the most general transformations satisfying all the requests and that corresponds to the set of infinitesimal transformations generating the BMS group. The special form of the transformations (2.78) derived from the requirement that these transformations must be diffeomorphisms of the compactified plane, $C \cup \{z = \infty\}$, into itself.

Hence, the transformations must have at least one pole, $z^\star$, and at least one zero, $z^{\star\star}$. The pole $z^\star$ corresponds to the point mapped to the north pole, $F(z^\star) = \infty$, while the zero corresponds to the point mapped to the south pole, $F(z^{\star\star})$. Consequently, the transformations have to be rational complex functions. The roots of the denominator and the numerator correspond to the points mapped to the north pole and south pole respectively.

Since the transformations have to be injective there must be only one point corresponding to the north pole and only one corresponding to the south pole. This implies that numerator and denominator are linear functions of $z$. With the request that the map is surjective we obtain the constraint $ad - bc \neq 0$ for the complex parameters $a,b,c,d$. We can rescale these parameters imposing $ad - bc = 1$.

Note that the transformations given by (2.78) are isomorphic to $SL(2,\mathbb{C})$, the group of the $2 \times 2$ complex special matrices with determinant 1. What is more $SL(2,\mathbb{C})$ is isomorphic to the orthochronous Lorentz group. In the last consideration we have showed that the non-abelian part of the BMS group is isomorphic to the orthochronous Lorentz transformations.

The second family of transformations involved in (2.74) are the so-called supertranslations, angle dependent translations of the retarded time, which assume, in complex coordinates, the form

$$u \mapsto u + \alpha(z,\bar{z}), \tag{2.79}$$

where $\alpha(z,\bar{z})$ is any smooth real function on the sphere and reproduces those that are the analogue of the Poincaré space-time translations. The functions $\alpha(z,\bar{z})$ are linear combinations of the following functions

$$1, \quad \frac{1-z\bar{z}}{1+z\bar{z}}, \quad \frac{z+\bar{z}}{1+z\bar{z}}, \quad \frac{i(z-\bar{z})}{1+z\bar{z}}.$$

These are nothing else that the spherical harmonics $\mathcal{Y}_0^0(\theta,\phi)$ and $\mathcal{Y}_m^1(\theta,\phi)$ in polar coordinates with $m = -1, 0, 1$. This is the most surprising result by Bondi et al., namely the statement that asymptotic symmetries enhance the Poincaré group to an infinite dimensional group with an infinite dimensional abelian normal subgroup of the supertranslations.

In the (2.5), we show the effect of a supertranslation in the case of a 3 spacetime dimensions.

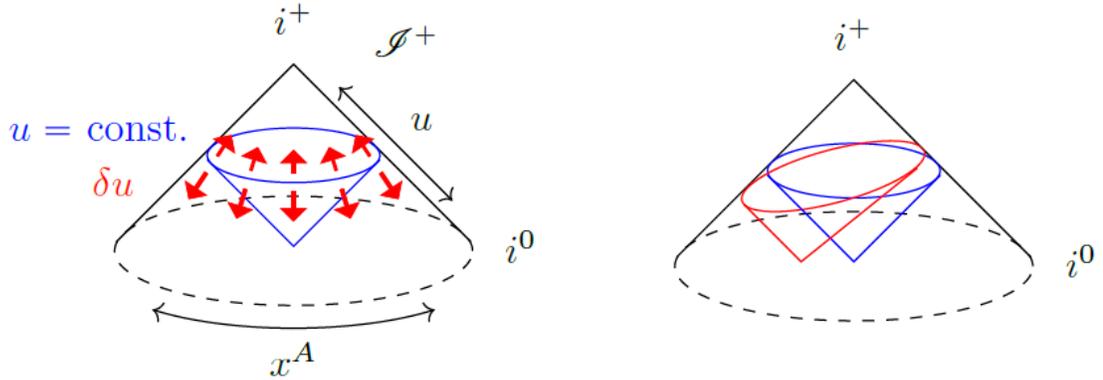

Figure 2.5: On the left the representation of the future null infinity cone $\mathscr{I} \times S^1$ in the case of three spacetime dimensions. Fixed $u$, it is defined a circle (the blue circle in the figure) on the cone representing the limit $r \to \infty$ of a future light-cone of some event. On the right, it is represented the effect of the supertranslation. The supertranslation has the effect of shifting the point from which the event originates and of deforming the circle [27]

## 2.4 Something about the Extended BMS group

In the previous section we have reviewed the original BMS group for which the conformal transformations are of type (2.78). However, there is no reason to impose such a restriction on the asymptotic conformal transformations. On the contrary, the restriction to conformal transformations (2.78) hides phisically relevant symmetries.

First, Banks in [30] suggests an enhancement of the global BMS group. Inspite of (2.78), one can "relax" the form of the conformal transformations requiring for the transformations to be generated by arbitrary conformal Killing vector fields on the celestial 2-sphere. The previous global conformal transformation (2.78) can be replaced by a local conformal transformation

$$Y^z = f(z) + O(1/r) \tag{2.80}$$

where $f(z)$ is a meromorphic function. The same still remain valid for $Y^{\bar{z}}$.

These new conformal transformations have in general poles at certain points. Despite the singularities, the transformations (2.80) preserve the asymptotic behaviour of the metric, thus they can be classified as asymptotic symmetries. $Y^z$ and $Y^{\bar{z}}$ admit a Laurent expansion

$$Y^z = \sum_{n \in \mathbb{Z}} a_n z^{n+1} \partial_z, \quad Y^{\bar{z}} = \sum_{n \in \mathbb{Z}} \bar{a}_n \bar{z}^{n+1} \partial_{\bar{z}} \tag{2.81}$$

and a basis for this set of solutions is

$$l_m = -z^{m+1} \partial_z, \quad \bar{l}_m = -\bar{z}^{m+1} \partial_{\bar{z}} \tag{2.82}$$

where $m \in \mathbb{Z}$. These bases are infinite and contain the six vector fields which generate the invertible Möbius transformations and the new singular vector fields that do not define diffeomorphisms of $S^2$ into itself. The new singular transformations are called superrotations. We have briefly explained what the extended BMS group is for completeness, but it does not play any role in this dissertation. For more detailed about the BMS group see, i.e., [24].

# Chapter 3

# The Carroll group

> "Well, in our country," said Alice, still painting a little, "you'd generally get to somewhere else if you run very fast for a long time, as we've been doing."
> "A slow sort of country!" said the Queen. "Now, here, you see, it takes all the running you can do, to keep in the same place. If you want to get somewhere else, you must run at least twice as fast as that!"

I steal the same dialogue reported in [32] betweeen Alice and her main antagonist, the Red Queen, from *Through the Looking-Glass* [33], the sequel of "Alice in Wonderland", to introduce what is called the Carroll group. The name is referred to the pseudonym of the author of this novel, Lewis Carroll. While the standard contraction of the Poincaré group yields the Galilei group [34], the Carroll group was introduced by Lévy-Leblond [36] as a less well known limit of the Poincaré group [1].

We will see that two different limits of the Poincaré group exist. The former is the well known Galilean approximation arising in the limit of low velocities and large space intervals. The other is the Carrollian limit appearing in the case of low velocities and large time intervals. Thus, we have the emergence of a new isometry group: the Carroll group.

The peculiarity of this new group is the absence of causality. A Universe lacking in causality would seem like the world where the adventures of "Alice in Wonderland" are set. This is the reason why Lévy-Leblond gave to the group the name of the author of the novels aforementioned.

---
[1]We clarify the concept of contraction of a group in what follow.

Galilei and Carroll groups present an interesting type of "*duality*" concerning two different concepts of "time". We denote by $t$ and $s$ the absolute Newtonian time and the Carrollian time respectively. Galilei and Carroll appear as two distinct symmetry groups of two different $(d+1)$-dimensional *non-Minkowskian* spacetimes. The Carroll group would seem to appear as a pure mathematical curiosity, but it has an important role in the field of the Asymptotic Symmetry Groups (ASG). In fact its conformal extension has been shown to be isomorphic to the BMS group [4]. We will analize the isomorphism between the two groups in the next chapter. What is more the Carroll group finds application in the study of plane gravitational waves and the memory effect [37], [38], as well as in the area of the flat holography [35].

In this chapter we show how the Carroll group emerges and what its algebra is. One way to find it is by the contraction of the Poincaré group, as just mentioned and as was done originally by Lévy-Leblond. Another way is to see the Carroll group as the group of isometries of the *non-Minkowskian* Carroll manifold. From this point of view the Carroll group has the analogous role of the Poincaré group for Minkowskian spacetime or of the Gailean isometries for the Newton-Cartan manifold.

Alternatively, we can consider the Carroll group as a subgroup of the $(d+1,1)$-dimensional Poincaré group, denoted by $E(d+1,1)$. This is the anologous of finding the Bargmann structure of a spacetime. A Bargmann structure allows to study the motion of $N$ non-relativistic point particles, attracting according to the inverse square law, through equations for null geodesics in a $(3d+2)$-dimensional Lorentzian spacetime, which is Ricci flat and admits a covariantly constant null vector. For more details see [39], [40].

## 3.1 Carroll group as Contraction of the Poincaré group

### 3.1.1 The two limits of the Poincaré group

Starting from the Poincaré group, we can find the connection between this group and two other groups, Galilei and Carroll, through the notion of contraction of groups due to Inönuü and Wigner [34]. We will consider the Carroll contraction of the Poincaré group but, first, it is worth to show how the Galilean contraction works. In this way we can also compare and contrast the two different approximations.

Consider for simplicity a Minkowski spacetime in 1 + 1 dimensions. Then, the interval between two events is $(\Delta x, \Delta t)$, and under a pure Lorentz transformation with velocity $u$, we have a new spacetime interval that we denote by $(\Delta x', \Delta t')$, where
$$\begin{cases} \Delta x' = \frac{\Delta x + u \Delta t}{\sqrt{1-u^2}} \\ \Delta t' = \frac{\Delta t + u \Delta x}{\sqrt{1-u^2}}. \end{cases}$$

We obtain the Galilean approximation by considering $u \ll 1$, so we have:

$$\begin{cases} \Delta x' = \Delta x + u \Delta t \\ \Delta t' = \Delta t. \end{cases}$$

The condition $u \ll 1$ is not sufficient to assure the validity of the Galilean approximation. In order to have $u \Delta t$ negligible with respect to $\Delta t$, we must require
$$\frac{\Delta x}{\Delta t} \ll 1. \tag{3.1}$$

Thus the Galilean approximation is valid in the regime of low velocities, $u \ll 1$, and large time intervals, condition given in (3.1).

Having worked out the Galilean limit, we now turn to the Carrollian one. It retains the same low-velocity condition as the Galilean case but assumes large space intervals
$$\frac{\Delta x}{\Delta t} \gg 1. \tag{3.2}$$

In this limit the Lorentz transformations reduce to

$$\begin{cases} \Delta x' = \Delta x \\ \Delta t' = \Delta t + u \Delta x. \end{cases} \tag{3.3}$$

We can illustrate what happens to the light-cone of the Minkowski spacetime taking the two limits (3.1) and (3.2). Introducing the new variables $\tau$ and $\lambda$ for times and lenghts respectively, the condition (3.1) means $\frac{\tau}{\lambda} \gg 1$ and, thus, the light-cone falls into the plane $t = 0$ (Galilean approximation). This reflects the fact that all the events of the spacetime are separated by timelike intervals; in this sense the light-cone vanishes and it collapses to the $t = 0$ axis. On the contrary, from condition (3.2) we have $\frac{\tau}{\lambda} \ll 1$ and the light-cone approaches the axis $x = 0$. This new non-relativistic limit yields the so called Carroll group. What we have described is illustrated in Figure (3.1).

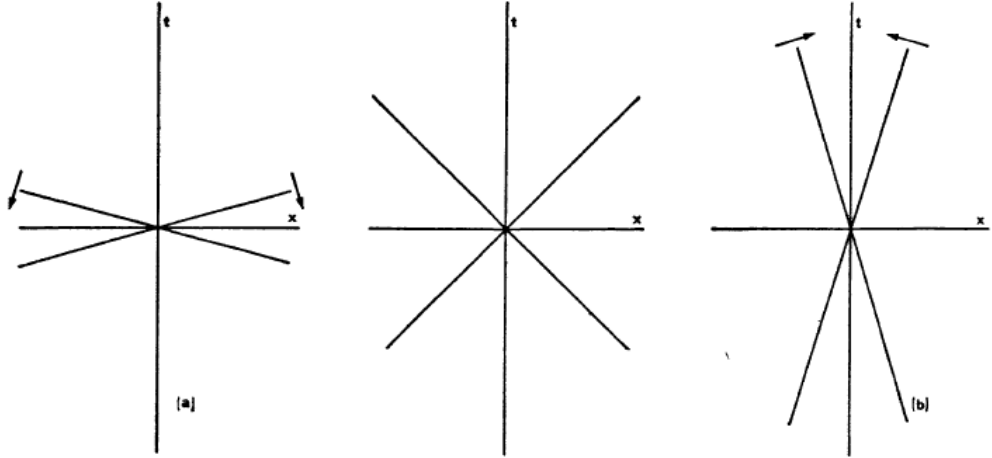

Figure 3.1: The figure in the middle is the light-cone in Minkowski spacetime. The figure on the left represents the Galilean limit of the light-cone structure and the figure on the right the Carroll limit.

### 3.1.2 The Carroll group and its Lie algebra

Let $(x_0, \mathbf{x})$ an event in Minkowski spacetime. Any Poincaré transformation can be devided in a spatial rotation R, a pure Lorentz transformation with velocity $\boldsymbol{\beta}$ and a spacetime translation $(l_0, \mathbf{l})$. Under the previous transformations the event $(x_0, \mathbf{x})$ becomes $(x'_0, \mathbf{x}')$, where

$$\begin{cases} x'_0 = \gamma(x_0 + \boldsymbol{\beta} \cdot R\mathbf{x}) + l_0 \\ \mathbf{x}' = R\mathbf{x} + \frac{\gamma^2}{\gamma+1}(\boldsymbol{\beta} \cdot R\mathbf{x}) + \gamma\boldsymbol{\beta} x_0 + \mathbf{l}, \end{cases} \quad (3.4)$$

with $\gamma = (1 - \beta^2)^{-\frac{1}{2}}$. Note that equation (3.4) has been written in a non-covariant form because the Carroll and Galilei limits, which we are going to take, are non-relativistic.

In order to obtain a general Galilean transformation starting from the Poincaré transformations, we replace the following change of coordinates

$$t = \frac{1}{c} x_0, \quad \mathbf{b} = c\boldsymbol{\beta}, \quad e = \frac{1}{c} l_0. \tag{3.5}$$

in (3.4) and take the limit $c \to \infty$. We obtain

$$\begin{cases} t' = t + e \\ \mathbf{x}' = R\mathbf{x} + \mathbf{b}t + \mathbf{l}. \end{cases}$$

On the other hand, setting

$$s = Cx_0, \quad \mathbf{b} = C\boldsymbol{\beta} \quad f = Cl_0, \tag{3.6}$$

in (3.4) and taking the limit $C \to \infty$ we obtain a general Carroll transformation

$$\begin{cases} s' = s + \mathbf{b} \cdot R\mathbf{x} + f \\ \mathbf{x}' = R\mathbf{x} + \mathbf{l}. \end{cases} \tag{3.7}$$

The Carroll group is the ten-parameter group of such transformations and it is, clearly, a subgroup of the group of general linear transformations $GL(4, \mathbb{R})$.

Because $C$ has the dimension of a velocity as $c$, it is clear that the Carrollian time $s$ has different dimension with respect to the absolute Galilean time $t$. The dimension of $s$ is $[s] = L^2 T^{-1}$, i.e., an *[action/mass]*.

From (3.7) it is evident that the Carroll group is generated by spatial rotations (R), pure Carroll transformations (**b**), space translations (**l**) and time translations ($f$). The composition law of the group is

$$(f', \mathbf{l}', \mathbf{b}', R') \cdot (f, \mathbf{l}, \mathbf{b}, R) = (f + f' + \mathbf{b}' \cdot R'\mathbf{l}, \, \mathbf{l}' + R'\mathbf{l}, \, \mathbf{b}' + R'\mathbf{b}, \, R'R). \tag{3.8}$$

The spacetime translations form a maximal abelian subgroup. Thus, the Carroll group appears as the semi-direct product of a group isomorphic to the isometries of a 3-dimensional Euclidean space, denoted by $\tilde{E}(3)$, and an abelian group of translations in 4 dimensions, $\mathbb{R}^4$

$$Carr(3+1) = \tilde{E}(3) \ltimes \mathbb{R}^4.$$

The same reasoning applies in an arbitrary $(d+1)$-dimensional Carrollian spacetime, $Carr(d+1) = \tilde{E}(d) \ltimes \mathbb{R}^{(d+1)}$. From the composition law (3.8) we can deduce the Carroll Lie algebra. Let $J_i, K_i, P_i$, with $i = 1, 2, 3$, the infinitesimal generators of rotations, boosts and spatial translations respectively and $P_0$ the generator of time translations.

In table (3.1) we compare the commutation rules defining Poincaré, Galilei and Carroll Lie algebras

| Poincaré | Galilei | Carroll |
|---|---|---|
| $[J_i, J_j] = i\epsilon_{ijk}J_k$ | $[J_i, J_j] = i\epsilon_{ijk}J_k$ | $[J_i, J_j] = i\epsilon_{ijk}J_k$ |
| $[J_i, K_j] = i\epsilon_{ijk}K_k$ | $[J_i, K_j] = i\epsilon_{ijk}K_k$ | $[J_i, K_j] = i\epsilon_{ijk}K_k$ |
| $[K_i, K_j] = -i\epsilon_{ijk}J_k$ | $[K_i, K_j] = 0$ | $[K_i, K_j] = 0$ |
| $[J_i, P_j] = i\epsilon_{ijk}P_k$ | $[J_i, P_j] = i\epsilon_{ijk}P_k$ | $[J_i, P_j] = i\epsilon_{ijk}P_k$ |
| $[K_i, P_j] = \delta_{ij}M$ | $[K_i, P_j] = 0$ | $[K_i, P_j] = i\delta_{ij}P_0$ |
| $[J_i, P_0] = 0$ | $[J_i, P_0] = 0$ | $[J_i, P_0] = 0$ |
| $[K_i, P_0] = iP_i$ | $[K_i, P_0] = iP_i$ | $[K_i, P_0] = 0$ |
| $[P_i, P_j] = 0$ | $[P_i, P_j] = 0$ | $[P_i, P_j] = 0$ |
| $[P_i, P_0] = 0$ | $[P_i, P_0] = 0$ | $[P_i, P_0] = 0$ |

Table 3.1: Comparison between Lie algebras.

If we carry out a redefinition of $K_i$ and $P_i$ as

$$K_i \to \epsilon K_i, \quad P_i \to \epsilon P_i$$

and we take the limit $\epsilon \to 0$ starting from the Poincaré Lie algebra we find the Galilean Lie algebra. On the other hand, replacing

$$K_i \to \eta K_i, \quad P_0 \to \eta P_0$$

and taking the limit $\eta \to 0$, we obtain the Carroll Lie algebra. Therefore, it is evident that the contraction of the Poincaré group with respect to the subgroup of spatial rotations and time translations yields the Galilei group. While, its contraction with respect to the spatial rotations and spatial translations yields to the Carroll group.

**Remarks**. A paradoxical aspect comes from the defining condition $\Delta x/\Delta t \gg 1$ of the Carrollian transformations. In Minkowski spacetime, two events separated by such an interval have no causal connection because they are separated by a spacelike distance $\Delta l^2 = -\Delta t^2 + \Delta x^2 \gg 0$. In the Carrollian limit the cones of the absolute past and of the absolute future collapse on the time axis; the region of the "absolute elsewhere" covers all the spacetime.

All the events are not causally related and the behavior of such a universe will be strange because of the freedom in the choice of the value of the time intervals. This freedom derives from the fact that it is the space interval to be an invariant in the Carroll theory, and not the time interval as in the Galilean theory.

## 3.2 Geometrical definition of the Carrol manifold

In the last section we have shown how to derive Galilei and Carroll groups as contractions of the Poincaré group. However, it is also possible to see Carroll manifolds as the dual of the Newton-Cartan manifolds [32]. Detailed discussions about the Newton-Cartan manifold (NC) and its symmetries are found in [41], [42], [43], [44].

### 3.2.1 Newton-Cartan manifold

We start recalling some results clearly explained in [44] about the Galilei (or Newton-Cartan structure). A *weak Newton-Cartan* structure is a triplet $(N, \gamma, \theta)$ formed from a $(d+1)$-dimensional manifold $N$ (in our case a $(3+1)$-dimensional), a twice symmetric contravariant positive tensor field $\gamma$ and a 1-form $\theta$ generating the kernel of $\gamma$. If we also define on the manifold a symmetric affine connection that parallel-transports $\gamma$ and $\theta$, we obtain the so-called *strong Newton-Cartan structure*. See, i.e., [45] for some notions on differential geometry used here.

The 1-form $\theta$ is closed: $d\theta = 0$ is the sufficient and necessary condition to guarantee a locally flat Newton-Cartan structure [44]. In fact, now, we are analyzing the standard flat Newton-Cartan structure. The kernel of $\theta$ gives rise to $d$-dimensional foliations each of which is endowed with a Riemannian structure due to $\gamma$. The absolute Newtonian time axis (1-dimensional) is done by the quotient [2] $K = N/ker\theta$.

---

[2]$K$ stays for "Kronos", the god of time and agriculture in the Greek mythology.

In 3 + 1 dimension, the symmetric contravariant vector field is for a flat NC structure

$$\gamma = \delta^{\alpha\beta} \frac{d}{dx^\alpha} \otimes \frac{d}{dx^\beta}; \tag{3.9}$$

moreover

$$N^4 = \mathbb{R}^3 \times \mathbb{R}, \quad \theta = dt, \quad \Gamma^k_{ij} = 0, \tag{3.10}$$

where $t$ is the absolute Galilean time coordinate, the integer indices $i, j, k$ range from 0 to 4 and the Christoffel symbols are zero as they have to be for a flat space-time.

The set of automorphisms of the NC structure forms the Galilei group, in 4 space-time dimensions pointed out by $Gal(3+1)$. $Gal(3+1)$ can be represented by the following matrices [46], [47]

$$a_{NC} = \begin{pmatrix} R & \mathbf{b} & \mathbf{l} \\ 0 & 1 & e \\ 0 & 0 & 1 \end{pmatrix}, \tag{3.11}$$

where $R \in O(3)$ represents the orthogonal transformation (rotations), $\mathbf{b}$ and $\mathbf{l} \in \mathbb{R}^3$ represent boosts and space translations and $e \in \mathbb{R}$ the time translations.

We can see that the Galilei Lie algebra is isomorphic to the Lie algebra of vector fields on the manifold $N$ whose expression can be found as follows.

Consider the matrix (3.11) acting on a 5-component vector $(\mathbf{x}, t, 1)$

$$\begin{pmatrix} R & \mathbf{b} & \mathbf{l} \\ 0 & 1 & e \\ 0 & 0 & 1 \end{pmatrix} \begin{pmatrix} \mathbf{x} \\ t \\ 1 \end{pmatrix} = \begin{pmatrix} R^1_i x^i + b^1 t + l^1 \\ R^2_i x^i + b^2 t + l^2 \\ R^3_i x^i + b^3 t + l^3 \\ t + e \\ 1 \end{pmatrix}, \tag{3.12}$$

where the vector on the right has been written in terms of the components of the matrix $R$ and the vectors $\mathbf{b}$ and $\mathbf{l}$.

For an infinitesimal transformation $R$ we can write

$$R = I_{3\times 3} + \epsilon \begin{pmatrix} 0 & \omega^1_2 & -\omega^1_3 \\ -\omega^1_2 & 0 & \omega^2_3 \\ \omega^1_3 & -\omega^2_3 & 0 \end{pmatrix} + O(\epsilon^2) = I_{3\times 3} + \epsilon i J + O(\epsilon^2), \tag{3.13}$$

with $I_{3\times 3}$ the identity matrix and $J$ the $3\times 3$ matrices of the generators of rotations around the three axes.

Let
$$b^i = \epsilon v^i, \quad l^i = \epsilon a^i, \quad e = \epsilon k, \tag{3.14}$$

then, by (3.14) and (3.13), the vector on the right hand side of (3.2.1) becomes

$$\begin{pmatrix} \mathbf{x} + \epsilon i J \mathbf{x} + \epsilon \mathbf{v} + \epsilon \mathbf{a} \\ t + \epsilon k \\ 1 \end{pmatrix}. \tag{3.15}$$

There will be a linear operator defined by

$$\chi = \xi^i(\mathbf{x}, t)\partial_i + \xi^0 \partial_0, \tag{3.16}$$

such that

$$\chi \begin{pmatrix} \mathbf{x} \\ t \\ 1 \end{pmatrix} = \begin{pmatrix} \xi \\ \epsilon \\ 0 \end{pmatrix}, \tag{3.17}$$

with $\epsilon = \xi^0$. By comparison with (3.15) we have

$$\xi^i = \begin{pmatrix} \omega_2^1 x^2 - \omega_3^1 x^3 + v^1 t + a^1 \\ \omega_1^2 x^1 + \omega_3^2 x^3 + v^2 t + a^2 \\ -\omega_1^3 x^1 + \omega_2^3 x^2 + v^3 t + a^3 \end{pmatrix}, \tag{3.18}$$

from which

$$\chi = (\omega_2^1 x^2 - \omega_3^1 x^3 + v^1 t + a^1)\partial_1 + \\ + (\omega_1^2 x^1 + \omega_3^2 x^3 + v^2 t + a^2)\partial_2 + \\ + (-\omega_1^3 x^1 + \omega_2^3 x^2 + v^3 t + a^3)\partial_3 + \epsilon \partial_t. \tag{3.19}$$

We can write the (3.19) in a more compact form

$$\chi = (\omega_\beta^\alpha x^\beta + v^\alpha t + a^\alpha)\partial_\alpha + \epsilon \partial_t. \tag{3.20}$$

This represents the full expression of the vector fields whose Lie algebra is isomorphic to the Lie algebra of the Galilei group, $\mathfrak{gal}(3+1)$.

### 3.2.2 Carroll manifolds

In analogy with the geometrical definition of the NC structure, we can state something similar for the Carrol manifold, following [32].

The Carroll structure can be identified by the quadruplet $(C, g, \zeta, \nabla)$. As previously, $C$ is the $(d+1)$-dimensional manifold ($3+1$ for us), $g$ the contravariant twice-symmetric tensor, positive defined, whose kernel is now generated by the vector field $\zeta$ and $\nabla$ is a symmetric affine connection that parallel-transports $g$ and $\zeta$. Note that, since the metric $g$ is degenerate, the connection $\nabla$ is not singled out by the pair $(g, \zeta)$.

The Carroll group is the set of diffeomorphisms that preserve the metric $g$, the vector field $\zeta$ and the connection $\nabla$. The Carroll structure is analogous to the NC structure apart from the substitution of the Galilean time coordinate $t$ with the Carrollian time coordinate $s$; in fact we have

$$C^{3+1} = \mathbb{R}^3 \times \mathbb{R}, \quad g = \delta_{\alpha\beta} \otimes dx^\alpha \otimes dx^\beta, \quad \zeta = \partial_s, \quad \Gamma^k_{ij} = 0. \tag{3.21}$$

From (3.21), it is clear the duality with the NC, the only differences are in the substitutions

$$t \to s \quad , \quad 1-\text{form } \theta \to \text{ vector field } \zeta. \tag{3.22}$$

In order to mantain this structure the Carroll lie algebra can be identified as the Lie algebra of the vector fields $\in C$ satisfying

$$\mathcal{L}_X g = 0, \quad \mathcal{L}_X \zeta = 0, \quad \mathcal{L}_X \nabla = 0. \tag{3.23}$$

The metric $g$ is invariant under the space and time translations as follows

$$x'^\alpha = x^\alpha, \quad s' = s + f(x^1, x^2, x^3) \tag{3.24}$$

with $\alpha = 1, 2, 3$ and $f$ a smooth function. Note that $f$ must be a constant because of the preservation of the affine connection $\nabla$. Since $g$ is invariant under space and time translations, its isometry group is infinite dimensional.

As for NC, we can put the automorphisms of the Carroll structure in a matrix form

$$a_C = \begin{pmatrix} R & 0 & \mathbf{1} \\ -\mathbf{b}^T R & 1 & f \\ 0 & 0 & 1 \end{pmatrix}, \tag{3.25}$$

again with $R \in O(3)$ orthogonal transformations, $\mathbf{b}$ and $\mathbf{l} \in \mathbb{R}^3$ boosts and space translations, $f \in \mathbb{R}$ time translations. $\mathbf{b}^T$ is the transposed vector of $\mathbf{b}$.

Following the same procedure used in the case on NC group we can find the explicit expression of the vector fields $X$. Again these vector fields are those for which the associated Lie algebra is isomorphic to the Carroll Lie algebra.

When $a$ acts on a vector $(\mathbf{x}, s, 1)$ we have

$$\begin{pmatrix} R & 0 & \mathbf{l} \\ -\mathbf{b}^T R & 1 & f \\ 0 & 0 & 1 \end{pmatrix} \begin{pmatrix} \mathbf{x} \\ s \\ 1 \end{pmatrix} = \begin{pmatrix} R\mathbf{x} + \mathbf{l} \\ -\mathbf{b}^T R \cdot \mathbf{x} + s + f \\ 1 \end{pmatrix}. \tag{3.26}$$

Recalling that $R$ is given by (3.13) and adopting the following changes of variables

$$l^i = \epsilon a^i, \quad b^{iT} = \epsilon v_i, \quad f = \epsilon \phi, \tag{3.27}$$

the vector on the right of (3.26) becomes

$$\begin{pmatrix} \mathbf{x} + \epsilon i J \mathbf{x} + \epsilon \mathbf{a} \\ -\epsilon \mathbf{v} \cdot \mathbf{x} - \epsilon^2 \mathbf{v} J \mathbf{x} + \epsilon \phi + s \\ 1 \end{pmatrix}. \tag{3.28}$$

Note that we can neglect the term of the second order in $\epsilon$ in agreement with (3.13). Thus, there will be a linear operator

$$X = \xi^i(\mathbf{x}, s) \partial_i + \xi^0 \partial_0 \tag{3.29}$$

such that

$$X \begin{pmatrix} \mathbf{x} \\ s \\ 1 \end{pmatrix} = \begin{pmatrix} \xi \\ \epsilon \\ 0 \end{pmatrix}. \tag{3.30}$$

By the comparison between (3.30) and (3.28) we have

$$\xi^i = \begin{pmatrix} \omega_2^1 x^2 - \omega_3^1 x^3 + a^1 \\ \omega_1^2 x^1 + \omega_3^2 x^3 + a^2 \\ -\omega_1^3 x^1 + \omega_2^3 x^2 + a^3 \end{pmatrix}, \tag{3.31}$$

from which
$$\xi^\alpha = \omega^\alpha_\beta x^\beta + a^\alpha . \tag{3.32}$$

Analogously we find
$$\xi^0 = \phi - v_\alpha x^\alpha . \tag{3.33}$$

Finally, the form of the vector fields searched is
$$X = (\omega^\alpha_\beta x^\beta + a^\alpha)\partial_\alpha + (\phi - v_\alpha x^\alpha)\partial_s . \tag{3.34}$$

### 3.2.3 Bargmann structure

The issue of this section is to introduce briefly what a Bargmann structure is, in order to show how it is possible to deduce Newton-Cartan and Carroll structures starting from a Bargmann one.

The main property of the Bargmann manifold is to have one extra dimension with respect to NC and Carroll and it is characterized by the triplet $(B, G, \zeta)$. $B$ stays for Bargmann and is a $d + 2$-dimensional manifold, $G$ is the metric associated with $B$ and it has signature $(d + 1, 1)$ and the vector $\zeta$ is a nowhere vanishing, complete, null vector, parallel-transported by the Levi-Civita connection

$$B = \mathbb{R}^d \times \mathbb{R} \times \mathbb{R}, \quad G = \sum_{A,B=1}^{d} \delta_{AB} dx^A \otimes dx^B + dt \otimes ds + ds \otimes dt, \quad \zeta = \frac{\partial}{\partial s} . \tag{3.35}$$

For a detailed discussion about the Bargmann structures see [40]. The vector $\zeta$ can be thought as a vertical vector and we can naivly visualiz the Bargmann space as in figure (3.2).

In fact, the vector $\zeta = \partial_s$ generates the vertical translations. Factoring out the Bargmann space with respect to $\zeta\mathbb{R}$ we find the Newton-Cartan structure (see figure (3.3)).

However, we can observe that the $(d + 1)$-dimensional section obtained for $t = const$ from $B$ is endowed with a Carroll structure $\forall t \in \mathbb{R}$. It is a 1-parameter family preserving the singular metric $\delta_{\alpha\beta} dx^\alpha dx^\beta$, see figure (3.4).

As in the NC and Carroll cases, we can define the matrix representation of the

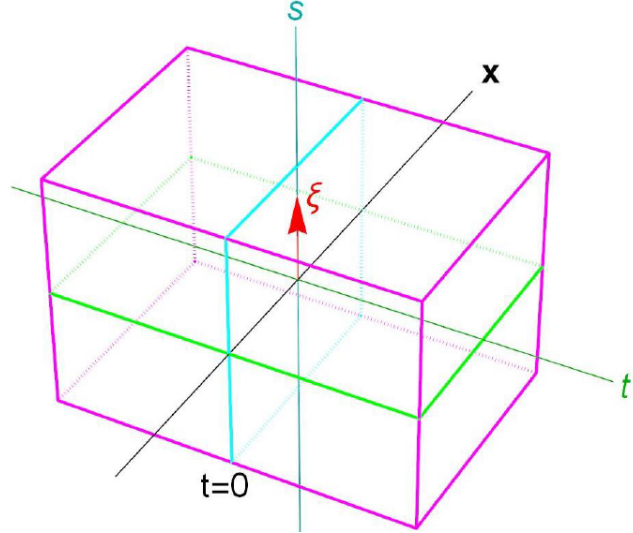

Figure 3.2: Bargmann space: a $(d, 1)$ manifold with a Lorentzian metric, defined by the coordinates $(\mathbf{x}, t, s)$ and the "vertical" vector $\zeta = \partial_s$ [48].

Bargmann group. We have

$$a_B = \begin{pmatrix} R & \mathbf{b} & 0 & \mathbf{l} \\ 0 & 1 & 0 & e \\ -\mathbf{b}^T R & -\frac{1}{2}\mathbf{b}^2 & 1 & f \\ 0 & 0 & 0 & 1 \end{pmatrix}, \tag{3.36}$$

with $R \in O(d)$ the orthogonal transformations, $\mathbf{b}$ and $\mathbf{l} \in \mathbb{R}^d$ boosts and space translations, $f$ and $e \in \mathbb{R}$ the Carrollian time and Galilean time translations. From (3.36) we can deduce the vector fields $Y$ whose Lie algebra is isomorphic to the Lie algebra of the Bargmann group.

Then, as previously

$$\begin{pmatrix} R & \mathbf{b} & 0 & \mathbf{l} \\ 0 & 1 & 0 & e \\ -\mathbf{b}^T R & -\frac{1}{2}\mathbf{b}^2 & 1 & f \\ 0 & 0 & 0 & 1 \end{pmatrix} \begin{pmatrix} \mathbf{x} \\ t \\ s \\ 1 \end{pmatrix} = \begin{pmatrix} R\mathbf{x} + \mathbf{b}t + \mathbf{l} \\ t + e \\ -\mathbf{b}^T R \cdot \mathbf{x} - \frac{1}{2}\mathbf{b}^2 + s + f \\ 1 \end{pmatrix}. \tag{3.37}$$

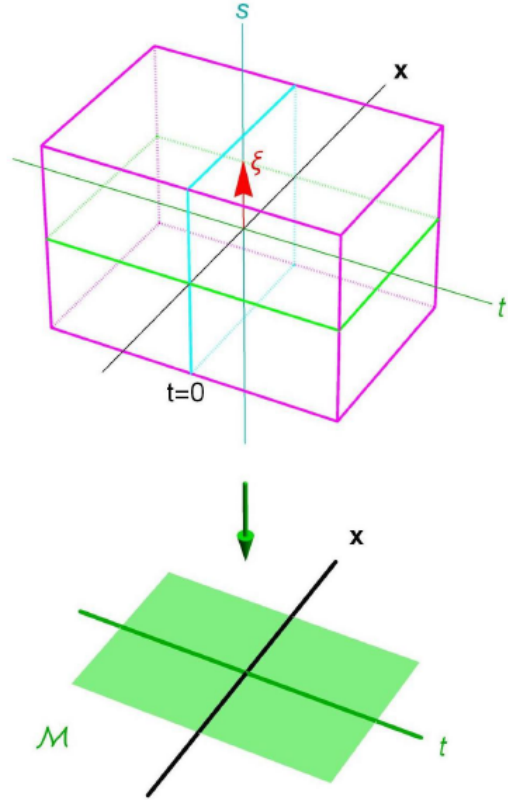

Figure 3.3: Newton-Cartan structure as the $d + 1$-dimensional quotient $N = B/\mathbb{R}\zeta$ [48].

Putting in this last vector (3.13), we have

$$\begin{pmatrix} \mathbf{x} + \epsilon i J \mathbf{x} + \mathbf{b}t + \mathbf{l} \\ t + e \\ -\mathbf{b}^T \mathbf{x} - \mathbf{b}^T \epsilon i J \mathbf{x} - \tfrac{1}{2}\mathbf{b}^2 + s + f \\ 1 \end{pmatrix}. \tag{3.38}$$

Let

$$b^i = \epsilon v^i, \quad l^i = \epsilon a^i, \quad e = \epsilon k, \quad f = \epsilon \phi, \quad b^{iT} = \epsilon v_i, \tag{3.39}$$

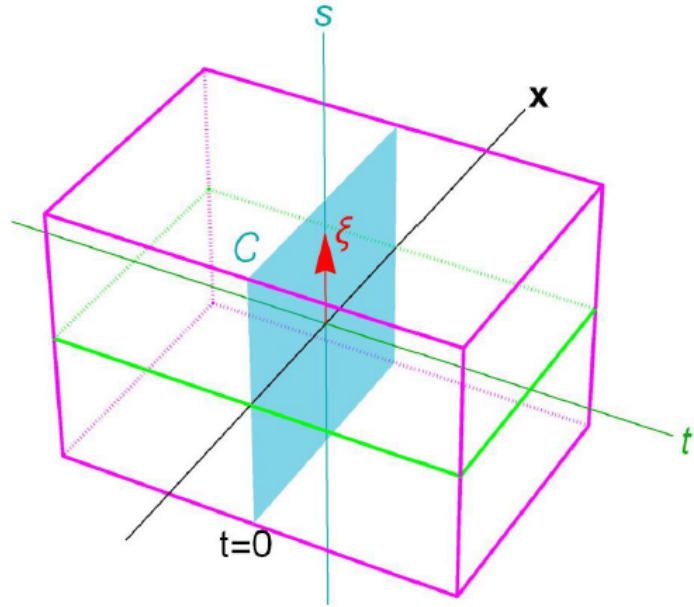

Figure 3.4: Carroll structure from the Bargmann space [48].

then (3.38) becomes

$$\begin{pmatrix} \mathbf{x} + \epsilon i J \mathbf{x} + \epsilon \mathbf{v} t + \epsilon \mathbf{a} \\ t + \epsilon k \\ -\epsilon \mathbf{v} \cdot \mathbf{x} - \epsilon^2 \mathbf{v} i J \mathbf{x} - \tfrac{1}{2}\epsilon^2 \mathbf{v}^2 t + s + \epsilon \phi \\ 1 \end{pmatrix}. \tag{3.40}$$

As usual, there will be a linear operator

$$Y = \xi^i(\mathbf{x}, s, t)\partial_i + \xi^0 \partial_0 \tag{3.41}$$

such that

$$Y \begin{pmatrix} \mathbf{x} \\ t \\ s \\ 1 \end{pmatrix} = \begin{pmatrix} \xi \\ \epsilon \\ \epsilon \\ 0 \end{pmatrix}. \tag{3.42}$$

Finally, by comparison, we find

$$Y = (\omega^\alpha_\beta x^\beta + v^\alpha t + a^\alpha)\partial_\alpha + k\partial_t + (\phi - v_\alpha x^\alpha)\partial_s . \tag{3.43}$$

The vector fields $Y$ generate the infinitesimal Bargmann transformation. Note that they are exactly the composition of the $\chi$, the vector fields generating the infinitesimal Galilean transformations, and $X$, the vector fields related to the Carroll transformations. This is in agreement with what we have stated about the Carroll and Galilei structures as two particular reductions of a Bargmann structure.

# Chapter 4

# The Carroll and BMS groups

Until now we have analyzed the standard BMS and Carroll groups. Moreover, a strict relation exists between the two groups. Specifically, we will see that the conformal extension of the Carroll group shows the presence of the supertranslations and, in particular, that the conformal Carroll transformations of level $N$ are exactly the same of the BMS transformations [4]. We clarify the concepts and the terminology in what follows.

## 4.1 The Conformal Carroll group

We would now investigate the conformal extension of the Carroll group. A conformal extension of the Galilei group also exists [49] [50], but for the purposes of our dissertation we will omit its analysis.

Given the standard flat Carroll structure $(C, g, \zeta)$, with $C = \mathbb{R}^d \times \mathbb{R}$, $g = \delta_{\alpha\beta} dx^\alpha dx^\beta$ and $\zeta = \frac{\partial}{\partial s}$, the conformal Carroll transformation $a$ of level $N$ (with $N$ an integer) is one for which the tensor $g \otimes \zeta^{\otimes N} = \Lambda$ is preserved, namely

$$a^*(\Lambda) = \Lambda. \tag{4.1}$$

The level $N$ means that $\zeta$ is taken $N$ times, so that $\zeta^{\otimes N} = \zeta \otimes \zeta \otimes ... \otimes \zeta$. This transformation satisfies

$$a^* g = \Omega^2 g \quad , \quad a_* \zeta = \Omega^{-\frac{2}{N}} \zeta. \tag{4.2}$$

The expression (4.1) can be put in terms of the Lie derivative of $\Lambda$ with respect to any of the vector field $X$ spanning the Lie algebra of the infinitesimal conformal Carroll transformations, denoted by $\mathfrak{ccarr}_N(C, g, \zeta)$, namely

$$\mathcal{L}_X \Lambda = 0, \tag{4.3}$$

From the (4.3) derives

$$\begin{aligned}\mathcal{L}_X g &= \lambda g \\ \mathcal{L}_X \zeta &= \mu \zeta \quad \text{with} \quad \lambda + N\mu = 0\end{aligned} \tag{4.4}$$

for some function $\lambda$ on $C$. The vector fields $X$ are such that

$$\begin{cases} X = X^\alpha \frac{\partial}{\partial x^\alpha} + T \frac{\partial}{\partial s} \\ \partial_s X^\alpha = 0 \\ \partial_s T = -\mu. \end{cases} \tag{4.5}$$

Solving equations (4.5) and (4.3), we find the expression

$$\begin{aligned} X = &\left(\omega^\alpha_\beta x^\beta + a^\alpha + (\chi - 2\kappa_\beta x^\beta)x^\alpha + \kappa^\alpha x_\beta x^\beta\right)\frac{\partial}{\partial x^\alpha} + \\ &\left(\frac{2}{N}(\chi - 2\kappa_\beta x^\beta)s + F(x^\alpha)\right)\frac{\partial}{\partial s}. \end{aligned} \tag{4.6}$$

The first line derives from the solution of the first of the equations (4.5) and represents the infinitesimal generators of the conformal group (see, i.e., chapter 4 of [51]). Specifically, $\omega^\alpha_\beta x^\beta \frac{\partial}{\partial x^\alpha}$ is the generator of rotations ($\omega \in \mathfrak{so}(d)$), $a^\alpha \frac{\partial}{\partial x^\alpha}$, $(\chi - 2\kappa_\beta x^\beta)x^\alpha \frac{\partial}{\partial x^\alpha}$ and $\kappa^\alpha x_\beta x^\beta \frac{\partial}{\partial x^\alpha}$ are the generators of translations, dilations and special conformal transformations respectively ($a, \kappa \in \mathbb{R}^d, \chi \in \mathbb{R}$).

The second line is obtained by the other two equations in (4.5). In particular from $\partial_s X^\alpha = 0$ we have a function $F(x^\alpha)$ depending only on the spatial coordinates $x^\alpha$ and from the last equation derives the fact that $T$ is linear in $s$. Precisely, $T = \frac{2}{N}(\chi - 2\kappa_\beta x^\beta)s$ with $T \in C^\infty(\mathbb{R}^d, \mathbb{R})$, where the conformal factor is expressed as

$$\lambda = 2(\chi - 2\kappa_\beta x^\beta). \tag{4.7}$$

Thus, the Carroll Lie algebra is infinite dimesional, since the function $F$ is an arbitrary function of the space variables. Note from (4.6) that, for $N = 2$, space and time transform under dilations in the same way.

We can obtain the non-conformal Carroll Lie algebra $\mathfrak{carr}(d+1)$ by choosing $\chi = \kappa = 0$ and $F = \phi - v_\alpha x^\alpha$, in agreement with the results and notation of the previous chapter. We have already seen that $\mathfrak{carr}(d+1)$ is spanned by the vector fields given by

$$X = (\omega^\alpha_\beta x^\beta + a^\alpha)\partial_\alpha + (\phi - v_\alpha x^\alpha)\partial_s \,. \tag{4.8}$$

Until now, we have considered the Carroll structure with the manifold $C = \mathbb{R}^d \times \mathbb{R}$. However, we can consider a more general Carroll structure given by

$$C = \Sigma \times \mathbb{R} \quad , \quad g = \tilde{g}_{\alpha\beta} dx^\alpha dx^\beta \quad , \quad \zeta = \frac{\partial}{\partial s} \quad , \quad \Gamma^C_{AB} = \tilde{\Gamma}^C_{AB} \,, \tag{4.9}$$

where $\Sigma$ is an hypersurface with Riemannian metric $\tilde{g}$ and $\Gamma^C_{AB} = \tilde{\Gamma}^C_{AB}$ are the non-vanishing components of the Carroll connection in this general case; precisely $\tilde{\Gamma}^C_{AB}$ is the Levi-Civita connection of $\tilde{g}$.

The vector field that span the general conformal Carroll Lie algebra $\mathfrak{ccarr}_N(C, g, \zeta)$ is

$$X = Y + \left(\frac{\lambda}{N} + F(x^\alpha)\right)\frac{\partial}{\partial s} \tag{4.10}$$

with $F(x^\alpha)$ a real function on $\Sigma$ and $Y = Y^A(x)\frac{\partial}{\partial x^A}$ the conformal vector field of $(\Sigma, \tilde{g})$. Then we have $\mathcal{L}_Y \tilde{g} = \lambda \tilde{g}$ and hence $\lambda = \frac{2}{d}\tilde{\nabla}_A Y^A$.

The group spanned by (4.10) acts on $(x, s)$ giving $(x', s')$, where the new coordinates are

$$\begin{aligned} x' &= \varphi(x) \\ s' &= \Omega^{\frac{2}{N}}(x)[s + \alpha(x)] \,, \end{aligned} \tag{4.11}$$

where $\varphi$ is a conformal transformation on the manifold $(\Sigma, \tilde{g})$ and $\alpha \in C^\infty$.

From the coordinate transformations (4.11) we can deduce that the general conformal Carroll group $CCarr_N(C, g, \zeta)$ is the semidirect product of the conformal transformations on $(\Sigma, \tilde{g})$, $Conf(\Sigma, \tilde{g})$, with the superstranslations, $ST$, of the Carrollian time

$$CCarr_N(C, g, \zeta) \equiv Conf(\Sigma, \tilde{g}) \ltimes ST \,. \tag{4.12}$$

In (4.11) we can observe that the transformation law of $s$ is analogous to the transformation law of the Bondi retarded time. Starting from the general conformal Carroll group, the BMS group is obtained when $N = 2$. In that case the manifold is $\Sigma = S^2$ endowed with the metric of the sphere. Then, the conformal transformations on the manifold $\varphi$ are that of the 2-sphere. We have already discussed in chapter 2 that the conformal group of $S^2$ is isormophic to $PSL(2, \mathbb{C})$, thus the conformal Carroll group of level $N = 2$ is given by

$$CCarr_2(S^2 \times \mathbb{R}, g, \zeta) \equiv PSL(2, \mathbb{C}) \ltimes ST. \tag{4.13}$$

### 4.1.1 A Carroll manifold: the Light-Cone

The light-cone of a $d + 1$-dimensional Minkowski spacetime can be considered as an example of the Carroll manifolds. It is possible to identify the future light-cone missing of $i^+$ and $i^0$, which we indicate as $C$, with $\mathscr{I}^+$. The region $\mathscr{I}^+$ is characterized by the fact that a time direction, respect to which we can define the dynamics, does not exist. Therefore, $\mathscr{I}^+$ does not include events with causal connections. This fact remembers the non-causality typical of a Carroll manifold.

We will show that $C$ has a structure anologous to the Carroll structure, namely it is endowed with a symmetric covariant positive tensor field $g$, whose kernel is generated by a nowhere vanishing vector field $\zeta$.

Let $\mathbb{M}$ be a Minkowski spacetime $\mathbb{R}^{d,1}$ and $G = diag(1, ..., 1, -1)$ its metric. We can describe the future null vectors generating $C$ with the pairs $(\mathbf{z}, t) \in \mathbb{R}^d \times \mathbb{R}$ such that $t \equiv |\mathbf{z}| > 0$. The light-cone $C$ is endowed with a symmetric tensor $g$ inherited from the Minkowski metric $G$. Since $t$ is positive and represents a radial coordinate on $C$, we can define a unit vector in the $\mathbf{z}$ direction as $\hat{x} = \mathbf{z}/t$. In this way the tensor $g$ takes the form

$$g = t^2 \hat{x}^2 = t^2 \tilde{g} \tag{4.14}$$

where $\tilde{g}$ is the round metric on the celestial sphere $S^{d-1}$. Thanks to the description of the null vector we have done, it is clear the fact that the projection of $g$ on the celestial sphere $S^{d-1}$ gives rise to a conformal class of metrics, denoted by $[g]$. Therefore, for each $t = const$, a map between the light-cone, defined by a precise value of $t$, and the corresponding celestial sphere, for that $t$, exists. We denote this map with $h$. Thus, each of those light-cones is generated by a vector field $\zeta = t\partial_t = \partial_s$, where $s = \log t$. The vector field $\zeta$ spans the kernel of the metric $g$, namely it is the element on $C$ such that $h(\zeta) = 0_{S^{d-1}}$.

This is in agreement with the fact that the metrics on the celestial spheres are degenerate [1]. By virtue of the considerations just done, we have shown that $C = \mathscr{I}^+$ is intrinsically a Carroll manifold. In our discussion, we have deliberately labeled the time coordinate with $s$ in order to underline the identification of the light-cone $C$ with a Carroll manifold.

The second step is to demonstrate that a connection compatible with the metric $g$ does not exist. In order to see this fact, we can consider a coordinate system $(x^A, s)$ on $C$ such that

$$g_{AB} = e^{2s}\tilde{g}_{AB}, \tag{4.15}$$

with $g_{As} = g_{ss} = 0$; here $e^{2s}$ represents a particular choice of the conformal factor. A symmetric affine connection $\nabla$ on $C$ should satisfy both

$$\nabla_s g_{AB} = 0 \quad , \quad \nabla_A g_{sB} = 0, \tag{4.16}$$

where $\nabla_k$ represents the covariant derivative and it is such that

$$\nabla_k g_{ij} = \partial_k g_{ij} - \Gamma^h_{kj} g_{hi} - \Gamma^h_{ik} g_{jh}, \tag{4.17}$$

with $\Gamma^k_{ij}$ the affine connection. The equations (4.16) are in contrast because of the presence in $\nabla_s g_{AB} = 0$ of the term $\partial_s g_{AB}$, which is not present in $\nabla_A g_{sB} = 0$.

Following the same procedure used in the previous section, we have for the future light-cone the following relations

$$\mathcal{L}_X g = \lambda g, \tag{4.18}$$

$$\mathcal{L}_Y \tilde{g} = (\lambda - 2X^s)\tilde{g}, \tag{4.19}$$

$$\mathcal{L}_X \zeta = -\frac{\lambda}{N}\zeta. \tag{4.20}$$

Starting from these conditions, in particular from the last equation, we can deduce that the conformal Carroll group of $C$ in $3 + 1$ dimensions, namely for $N = 2$ and $S^d = S^2$, is again

$$CCarr_C(S^2 \times \mathbb{R}, g, \zeta) \equiv PSL(2, \mathbb{C}) \ltimes ST. \tag{4.21}$$

---

[1] We can think, i.e., at the representative metric of $S^2$, $ds^2 = 0 \times du^2 + d\theta^2 + \sin^2\theta d\phi^2$, where $u$ is the retarded time.

We can fix the supertranslations by the choice $\lambda = 0$, from which

$$ST = X^s = -\frac{1}{2}(\mathcal{L}_Y \tilde{g})/\det \tilde{g}. \tag{4.22}$$

Thus, the time-part is subject to supertranslations, while the space-part is characterized by conformal invariance. Finally, the Carroll isometries of the light-cone, in $d$ dimension, span the conformal group $O(d,1)$ of the celestial sphere $S^d$ with the supertranslations fixed by the choice $\lambda = 0$.

## 4.2 Plane gravitational waves and Carroll symmetry

The Carroll group finds a physical application in the study of plane gravitational waves. In particular, we will show that the Carroll group without rotations represents the 5-dimensional isometry group of a gravitational plane wave [14].

A gravitational plane wave is a special class of a vacuum pp-wave spacetime, an important family of exact solutions of Einstein's equation, and may be defined in terms of Brinkmann coordinates by the Lorentz metric associated to the 4-manifold [52]

$$g_B = \delta_{ij} dX^i dX^j + 2dUdV + K_{ij}(U) X^i X^j dU^2, \tag{4.23}$$

Here we are following the notation used in [14] [53]. The $2 \times 2$ matrix $K_{ij}$, whose general form is

$$K_{ij}(U)X^i X^j = \frac{1}{2} A_+(U)\left((X^1)^2 + (X^2)^2\right) + A_\times X^1 X^2, \tag{4.24}$$

is traceless and symmetric and characterizes the profile of the wave. The quantities $A_+$ and $A_\times$ are the amplitudes of the two polaritazion states, usually labelled by + and ×. The term $K_{ij}(U)X^i X^j$ represents a quadratic potential. The coordinates $U$ and $V$ are light-cone ones and $\mathbf{X} = (X^1, X^2)$ parametrizes the transverse plane which carries the flat Euclidean metric $d\mathbf{X}^2 = \delta_{ij} dX^i dX^j$. Since the only non-vanishing components of the Riemann tensor are

$$R_{iUjU}(U) = -K_{ij}(U), \tag{4.25}$$

the metric is Ricci-flat. Hence, in the four dimensional spacetime the (4.23) represents the general form of a Ricci-flat Brinkmann metric.

In order to identify the transformations that leave invariant the (4.23) it is convenient to make a change of coordinates

$$(\mathbf{X}, U, V) \to (\mathbf{x}, u, v) \tag{4.26}$$

such that

$$\mathbf{X} = P(u)\mathbf{x}, \quad U = u, \quad V = v - \frac{1}{4}\mathbf{x} \cdot \dot{a}(u)\mathbf{x}, \tag{4.27}$$

where $a(u) = P^{\dagger}(u)P(u)$, with $P(u)$ a $2 \times 2$ non-singular matrix which satisfies

$$\ddot{P} = KP; \tag{4.28}$$

the dot stands for the derivative with respect to $u$. For the second-order ODE (4.28) we have $P^{\dagger}\dot{P} - \dot{P}^{\dagger}P = const$. Without loss of genarility, it is possible to choose the initial values of $\dot{P}$ and $P$ such that the constant vanishes

$$P^{\dagger}\dot{P} - \dot{P}^{\dagger}P = 0. \tag{4.29}$$

Under this change of coordinates the metric (4.23) takes the form

$$g_{BJR} = a_{ij}(u)dx^i dx^j + 2dudv. \tag{4.30}$$

The metric (4.30) is a plane wave in Baldwin-Jeffery-Rosen (BJR) coordinates. The relation between the two matrices appearing in (4.23) and (4.30), $K$ and $a$, is

$$K = \frac{1}{2}P\left(\dot{b} + \frac{1}{2}\right)P^{-1}, \quad b \equiv a^{-1}\dot{a}. \tag{4.31}$$

The non zero components of the Riemann tensor in BJR coordinates are

$$R_{uiuj} = -\frac{1}{2}\left(\ddot{a} - \frac{1}{2}\dot{a}a^{-1}\dot{a}\right), \tag{4.32}$$

yielding the non-zero components of the Ricci tensor is

$$R_{uu} = -\frac{1}{2}Tr\left(\dot{b} + \frac{1}{2}b^2\right). \tag{4.33}$$

In order to have a flat metric one has to solve $R_{uiuj} = 0$ with initial conditions

(following [53])
$$\begin{cases} a_0 = a(u_0) \\ \dot{a}_0 = \dot{a}(u_0) \end{cases}.$$

The solution is
$$a(u) = \left(a_0 + \frac{1}{2}(u - u_0)\dot{a}_0\right) a_0^{-1} \left(a_0 + \frac{1}{2}(u - u_0)\dot{a}_0\right) . \qquad (4.34)$$

Defining the matrix $c_0 = \frac{1}{2} a_0^{\frac{1}{2}} \dot{a}_0 a_0^{-\frac{1}{2}}$, equation (4.34) becomes
$$a(u) = a_0^{\frac{1}{2}} \left(\mathbf{1} + (u - u_0)c_0\right)^2 a_0^{\frac{1}{2}} . \qquad (4.35)$$

Making a change of coordinates from $(\mathbf{x}, u, v)$ to $(\tilde{\mathbf{x}}, \tilde{u}, \tilde{v})$ defined by:
$$\begin{cases} \tilde{\mathbf{x}} = (\mathbf{1} + (u - u_0)c_0) a_0^{\frac{1}{2}} \mathbf{x} \\ \tilde{u} = u \\ \tilde{v} = v - \frac{1}{2}\mathbf{x} \cdot \left(a_0^{\frac{1}{2}} c_0 (\mathbf{1} + (u - u_0)c_0) a_0^{\frac{1}{2}} \mathbf{x}\right) \end{cases}$$

whose inverse is
$$\begin{cases} \mathbf{x} = a_0^{-\frac{1}{2}} (\mathbf{1} + (u - u_0)c_0)^{-1} \tilde{\mathbf{x}} \\ \tilde{u} = u \\ v = \tilde{v} + \frac{1}{2}\tilde{\mathbf{x}} \cdot \left(c_0 (\mathbf{1} + (u - u_0)c_0)^{-1} \tilde{\mathbf{x}}\right) \end{cases},$$

then, it is possible to put the metric (4.30) in the standard Minkowskian form in any flat region
$$g = d\mathbf{x} \cdot a(u) d\mathbf{x} + 2 du dv = d\tilde{\mathbf{x}} \cdot a(u) d\tilde{\mathbf{x}} + 2 d\tilde{u} d\tilde{v} . \qquad (4.36)$$

After the discussion about the coordinate transformations from Brinkmann to BJR, we come back to the problem of the isometries. Specifically, the isometry group of the metric (4.30) is a 5-dimensional Lie group. Souriau [54] proved that, in order to preserve the metric 4.30, the set $(\mathbf{x}, u, v)$ has to transform as follows

$$\begin{pmatrix} \mathbf{x} \\ u \\ v \end{pmatrix} \to \begin{pmatrix} \mathbf{x} + H(u)\mathbf{b} + \mathbf{l} \\ u \\ v - \mathbf{b} \cdot \mathbf{x} - \frac{1}{2}\mathbf{b} \cdot H(u)\mathbf{b} + f \end{pmatrix} \qquad (4.37)$$

where $\mathbf{l} \in \mathbb{R}^2$ and $f \in \mathbb{R}$ generate respectively the transverse-space and null translations along the $v$ coordinate, and the matrix $H$ is a generic primitive of $a^{-1}$

$$H(u) = \int_{u_0}^{u} a(t)^{-1} dt . \qquad (4.38)$$

In fact from the (4.37) we have the transformed quantities

$$\begin{cases} d\mathbf{x}' = d\mathbf{x} + \dot{H}(u)\mathbf{b} du \\ du' = du \\ dv' = dv - \mathbf{b} \cdot d\mathbf{x} - \frac{1}{2}\mathbf{b} \cdot \dot{H}(u) du , \end{cases} \qquad (4.39)$$

which, substituted in (4.30), give

$$\begin{aligned} g'_{BJR} &= a_{ij}(u) \left( dx^i + \dot{H}(u) b^i du \right) \left( dx^j + \dot{H}(u) b^j du \right) + 2du \left( dv - \mathbf{b} \cdot d\mathbf{x} - \frac{1}{2}\mathbf{b} \cdot \dot{H}(u)\mathbf{b} du \right) \\ &= a_{ij}(u) dx^i dx^j + a_{ij}(u) \left( \dot{H}(u) b^j dx^i du + \dot{H}(u) b^i dx^j du \right) + a_{ij}(u) \left( \dot{H}(u) \right)^2 b^i b^j du^2 + \\ &\quad + 2du dv - 2du \mathbf{b} \cdot d\mathbf{x} - \mathbf{b} \cdot \dot{H}(u)\mathbf{b} du^2 \\ &= a_{ij}(u) dx^i dx^j + 2du dv = g_{BJR} . \end{aligned} \qquad (4.40)$$

The last line is easily obtained by virtue of the (4.38).

What is more, we can deduce from the (4.37) the group law

$$(\mathbf{b}, \mathbf{l}, f) \cdot (\mathbf{b}', \mathbf{l}', f') = (\mathbf{b} + \mathbf{b}', \mathbf{l} + \mathbf{l}', f + f' - \mathbf{b} \cdot \mathbf{l}') . \qquad (4.41)$$

From a quick comparison with the Carroll group law (3.8), we can note that (4.41) is precisely the same composition law with no rotations, i.e. $R = \mathbb{I}$. Therefore, the isometry group of our plane gravitational wave is represented by the following

matrix
$$\begin{pmatrix} \mathbb{I}_{2\times 2} & \mathbf{b} & 0 & \mathbf{l} \\ 0 & 1 & 0 & 0 \\ -\mathbf{b}^\dagger & -\frac{1}{2}\mathbf{b}\cdot H(u)\mathbf{b} & 1 & f \\ 0 & 0 & 1 & 0 \end{pmatrix} \tag{4.42}$$

As before, $\mathbf{b} \in \mathbb{R}^2$ represents the Carroll boost, $\mathbf{l} \in \mathbb{R}^2$ and $f \in \mathbb{R}$ are the space and time translations respectively. Thus, it is a normal subgroup of the Carroll group in $2+1$ dimensions, namely $Carr(2+1)/Carr \cong O(2)$. We find that the generators of the Lie algebra are

$$\begin{cases} K_i = H_{ij}\partial_{x_j} + H_{ij}\partial_{x_j} - x_i\partial_v \\ P_i = \partial_{x_1} \\ P_0 = \partial_v, \end{cases} \tag{4.43}$$

where $i, j = 1, 2$. Thus we have five generators, two originates the Carroll boosts, two the space translations and one the time translation. The only non vanishing Lie brackets of the generators (4.43) are

$$[K_i, P_j] = \delta_{ij} P_0. \tag{4.44}$$

Now, we would compute the expressions of the conserved quantities associated to the isometries and the geodesics. We start defining the action concerning the metric in BJR coordinates (4.30)

$$S = \int \mathscr{L} ds = \int \frac{1}{2}\left[a_{ij}\dot{x}_i\dot{x}_j + 2\dot{u}\dot{v}\right] ds; \tag{4.45}$$

the generalized momenta are

$$\begin{cases} p_i = \frac{\partial \mathscr{L}}{\partial \dot{x}^i} = a_{ij}\dot{x}^j \\ p_u = \frac{\partial \mathscr{L}}{\partial \dot{u}} = \dot{v} \\ p_v = \frac{\partial \mathscr{L}}{\partial \dot{v}} = \dot{u}. \end{cases} \tag{4.46}$$

This may be derived from the Euler-Lagrange equations. In order to find the expression of the generalized momenta following the action of the Carroll group, we can rewrite more explicitly the (4.37) as follows

$$\begin{pmatrix} x_1' \\ x_2' \\ u' \\ v' \end{pmatrix} = \begin{pmatrix} 0 \\ 0 \\ 0 \\ f \end{pmatrix} + \begin{pmatrix} l_1 \\ l_2 \\ 0 \\ 0 \end{pmatrix} + \begin{pmatrix} H_{11}b_1 \\ H_{21}b_1 \\ 0 \\ -b_1 x_1 - \frac{1}{2} H_{11} b_1^2 \end{pmatrix} + \begin{pmatrix} H_{22}b_2 \\ H_{12}b_2 \\ 0 \\ -b_2 x_2 - \frac{1}{2} H_{22} b_2^2 \end{pmatrix} = \phi_f + \phi_l + \phi_{b_1} + \phi_{b_2}, \quad (4.47)$$

and then

$$d\phi_f = \begin{pmatrix} 0 \\ 0 \\ 0 \\ 1 \end{pmatrix}, \quad d\phi_{l_1} = \begin{pmatrix} 1 \\ 0 \\ 0 \\ 0 \end{pmatrix}, \quad d\phi_{l_2} = \begin{pmatrix} 0 \\ 1 \\ 0 \\ 0 \end{pmatrix}, \quad d\phi_{b_1}|_{b_1=0} = \begin{pmatrix} H_{11} \\ H_{21} \\ 0 \\ -x_1 \end{pmatrix}, \quad d\phi_{b_2}|_{b_2=0} = \begin{pmatrix} H_{22} \\ H_{12} \\ 0 \\ -x_2 \end{pmatrix}. \quad (4.48)$$

The transformed generalized momenta are given by

$$p_\alpha = \sum_i p_i d\phi_\alpha^i. \quad (4.49)$$

Thus, explicitly for $p_f$

$$p_f = (a_{1j} \dot{x}^j \; a_{2j} \dot{x}^j \; 0 \; \dot{u}) \cdot \begin{pmatrix} 0 \\ 0 \\ 0 \\ 1 \end{pmatrix} = \dot{u}. \quad (4.50)$$

Analogously for the others $p_\alpha$, we find

$$\begin{cases} p_{l_1} = a_{1j} \dot{x}^j \\ p_{l_2} = a_{j} \dot{x}^j \\ p_{b_1} = H_{11} a_{1j} \dot{x}^j + H_{21} a_{2j} \dot{x}^j - \dot{u} x_1 \\ p_{b_2} = H_{22} a_{2j} \dot{x}^j + H_{21} a_{1j} \dot{x}^j - \dot{u} x_2 \end{cases}, \quad (4.51)$$

namely, in a more compact way

$$\begin{cases} p_l = a\dot{\mathbf{x}} \\ p_{\mathbf{b}} = Hp_l - \dot{u}\mathbf{x} \end{cases}. \quad (4.52)$$

These quantities can be interpreted as the linear momentum and the boost-momentum respectively. Note that we have used $H$ instead of $H(u)$ in order to have a less awkward notation.

Let us $\mathbf{q} = (\mathbf{x}, u, v)$ and $\mathbf{p} = (a\dot{\mathbf{x}}, \dot{v}, \dot{u})$ be the generalized coordinates and momenta. From the first of the (4.46) we have $\dot{\mathbf{x}} = a^{-1}\mathbf{p}_1$. Then, the hamiltonian is

$$\begin{aligned}
\mathscr{H} &= p_i \dot{q}_i - \mathscr{L} = p_1(a^{-1}p_1)_1 + p_2(a^{-1}p_1)_2 + p_u \dot{u} + p_v \dot{v} - \mathscr{L} \\
&= p_1(a^{-1})_{1j}p_j + p_2(a^{-1})_{2j}p_j + p_u p_v + p_v p_u - \frac{1}{2}\left[a_{ij}(a^{-1})_{ik}p_k(a^{-1})_{jl}p_l + 2p_u p_v\right] \\
&= \mathbf{p}_1 \cdot a^{-1}\mathbf{p}_1 + 2p_u p_v - p_u p_v - \frac{1}{2}(a^{-1})_{ik}p_k \delta_{il}p_l \\
&= \frac{1}{2}\mathbf{p}_1 \cdot a^{-1}\mathbf{p}_1 + p_u p_v.
\end{aligned} \qquad (4.53)$$

Now, we can calculate the Poisson brackets

$$\begin{cases}
\dot{\mathbf{x}} = \{\mathbf{x}, \mathscr{H}\} = a^{-1}\mathbf{p}_1 \\
\dot{\mathbf{p}}_1 = \{\mathbf{p}_1, \mathscr{H}\} = 0 \\
\dot{u} = \{u, \mathscr{H}\} = p_v \\
\dot{v} = \{v, \mathscr{H}\} = p_u \\
\dot{p}_u = \{p_u, \mathscr{H}\} = \frac{1}{2}\mathbf{p}_1 \cdot a^{-1}\dot{a}a^{-1}\mathbf{p}_1 \\
\dot{p}_v = \{p_v, \mathscr{H}\} = 0,
\end{cases} \qquad (4.54)$$

from which

$$\mathbf{p}_1 = const = a\dot{\mathbf{x}}, \qquad (4.55)$$

the conservation of the linear momentum, and

$$p_v = const = \dot{u} \qquad (4.56)$$

which we can interpret as the mass and it is exactly equal to the unity, since we have choosen as affine parameter the coordinate $u$ itself. From the (4.54) we have a second order differential equation for $v$

$$\ddot{v} = \dot{p}_u = \frac{1}{2}\mathbf{p}_1 \cdot a^{-1}\dot{a}a^{-1}\mathbf{p}_1; \qquad (4.57)$$

then, integrating we have

$$\dot{v} = -\frac{1}{2}p_1 \cdot a^{-1} p_1 + W, \tag{4.58}$$

and finally, using also the (4.38), we find

$$v = -\frac{1}{2}p_1 \cdot Hp_1 + Wu + d, \tag{4.59}$$

which is a constant from the (4.54); the quantity $d$ is an integration constant. The rewrite the (4.55) and (4.59) including explicitly the dependence from the affine parameter $u$

$$\begin{cases} \mathbf{x}(u) = H(u)p_1 + x_0 \\ v(u) = -\frac{1}{2}p_1 H(u) p_1 + Wu + d; \end{cases} \tag{4.60}$$

these represent the geodesic equation in BJR coordinates. Since we have choosen $\dot{u} = 1$ and by virtue of the second of the (4.52), we can identify the integration constant $x_0$ with $p_\mathbf{b}$, thus

$$x(u) = H(u)p_1 - p_\mathbf{b}. \tag{4.61}$$

We have seen that three conserved exist, $\mathbf{p_l}$, $\mathbf{p_b}$ and $p_v$, which transform under the action of the Carroll group as

$$\begin{pmatrix} p_1 \\ p_\mathbf{b} \\ W \\ d \end{pmatrix} \rightarrow \begin{pmatrix} p_1 + 1 \\ p_\mathbf{b} + \mathbf{b} \\ W \\ d + f - p_1 \cdot p_\mathbf{b} \end{pmatrix} \tag{4.62}$$

It is evident the possibility to bring to zero the constants $p_1$, $p_\mathbf{b}$ and $d$, for the group action. Thus, the two geodesics become

$$\begin{aligned} \mathbf{x}(u) &= 0 \\ v(u) &= Wu. \end{aligned} \tag{4.63}$$

We can identify this $W$ as an etra conserved quantity, namely the *kinetic energy*. In fact, coming back to the Lagrangian defined in (4.45), $\mathscr{L} = \frac{1}{2}a_{ij}\dot{x}_i\dot{x}_j + 2\dot{u}\dot{v}$, we have for the kinetic energy, namely $W$ itself,

$$W = \frac{1}{2}g_{\mu\nu}\dot{x}^\mu \dot{x}^\nu = \frac{1}{2}\dot{\mathbf{x}} \cdot a(u)\dot{\mathbf{x}} + \dot{v}. \tag{4.64}$$

The geodesics are timelike/lightlike/spacelike, if $W$ are $< 0$, $= 0$, $> 0$, respectively. Therefore, for each sign of $W$ just one type of "vertical" geodesic exists, where for vertical we mean the fact that the geodesic has no component along $x$. In other words, we can state that a suitable action of the Carroll group has *straightened-out* the geodesic (see figure (4.1)); each geodesic can be affected by the same kind of transformation.

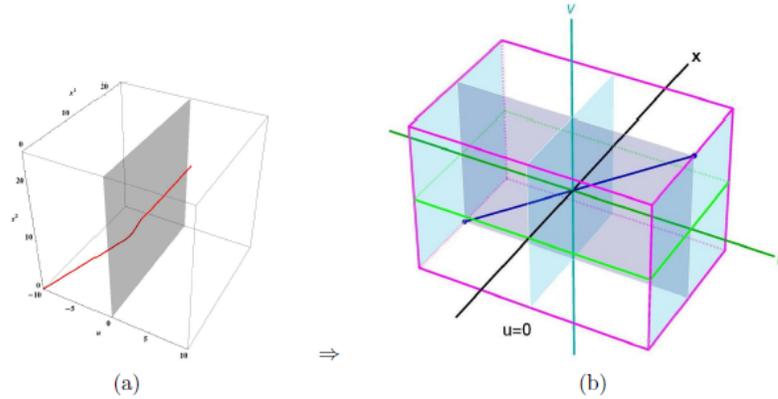

Figure 4.1: (a) The rapresentation of a generic geodesic. (b) The straightened geodesic following the action of the Carroll group [53].

### 4.2.1 BJR geodesics in the flat zones

In this subsection we would describe the geodesics in BJR coordinates in the flat zones. In order to clarify what we mean for flat zones, let us consider sandwich waves, i.e. gravitational waves which are flat outside the sandwich and non-flat in a particular interval $[u_i, u_f]$.

Thus, we can distinguish two different flat zones: the before zone, for $u < u_i$, and the after zone, for $u > u_f$. The flat after zone is non-equivalent to the before zone, since some information remain encoded by the effect of the passage of the gravitational wave.

Inside the region $[u_i, u_f]$, the Riemann tensor is not zero yet, but, since we are considering a system without matter, the Ricci tensor (4.33) is zero

$$Tr\left(\dot{b} + \frac{1}{2}b^2\right) = 0. \qquad (4.65)$$

The introduction of new variables defined as

$$\chi = (\det a)^{\frac{1}{4}} \quad and \quad \gamma = \chi^{-2} a \qquad (4.66)$$

allows to replace the (4.65) with the following second order differential equation

$$\ddot{\chi} + k^2(u)\chi = 0, \quad with \quad k^2(u) = \frac{1}{8} Tr((\gamma^{-1}\dot{\gamma})^2) \quad , \qquad (4.67)$$

where

$$\gamma(u) = \begin{pmatrix} \alpha(u) & \beta(u) \\ \beta(u) & \frac{(1+\beta(u)^2)}{\alpha(u)} \end{pmatrix}$$

is a unimodular symmetric 2×2 matrix which guarantees that the vacuum Einstein equations are satisfied. The choice of $\alpha(u)$ and $\beta(u)$ is arbitrary.

Since $(\gamma^{-1}\dot{\gamma})^2 > 0$, then $k^2$ also is positive. Thus, equation (4.67) describes an attractive oscillator with a time-dependent frequency. It follows that $\chi(u)$ is a concave function, $\ddot{\chi} < 0$, which implies the vanishing of $\chi$ for some $u_s < u_i$, $\chi(u_s) = 0$. The subscript $s$ on $u$ signals the fact that the metric (4.30) in BJR coordinates is singular for $u_s$.

In order to find the singularities in the BJR metric, it is convenient to study some physically relevant profile in Brinkmann coordinates [2] and solving numerically the equation (4.28) allows us to calculate the matrix $a$ and to plot $\chi(u)$ and $k^2(u)$ in (4.66) and (4.67).

A simple example (following [53]) is obtained choosing $a(u) = diag(a_{11}, a_{22})$, from which

$$k^2(u) = \frac{1}{16}\left(\frac{\dot{a}_{11}}{a_{11}} - \frac{\dot{a}_{22}}{a_{22}}\right)^2, \qquad (4.68)$$

---

[2]Note that the metric in Brinkmann coordinates does not present singularities.

and then the (4.67) becomes

$$\frac{\ddot{a}_{11}}{a_{11}} + \frac{\ddot{a}_{22}}{a_{22}} - \frac{1}{2}\left(\frac{\dot{a}_{11}^2}{a_{11}^2} + \frac{\dot{a}_{22}^2}{a_{22}^2}\right) = 0. \tag{4.69}$$

In terms of the $P$, the (4.69) is

$$\frac{\ddot{P}_{11}}{P_{11}} + \frac{\ddot{P}_{22}}{P_{22}} = 0 \tag{4.70}$$

which is indeed $Tr(K) = 0$ since $\dot{P}P^{-1} = K$. The numerical solution of the (4.70) can be visualized in figure (4.2).

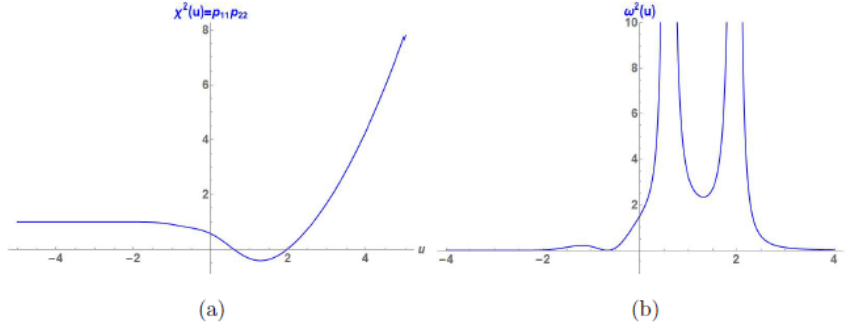

(a)      (b)

Figure 4.2: (Figures (a) and (b) are the plots of $\chi = (\det a)^{1/4}$ and $k^2(u)$, respectively. The results are obtained by solving (4.70) with the choice $A_+ = A_+^3$ [53].

Suppose, now, that $a(u) = 1$ in the before zone, i.e. for $u < u_i$. Noting that $u_s$ may lie in or outside the sandwich $[u_i, u_f]$, we can state that the BJR coordinates can be used only for $u < u_s$, which we will assume henceforth. Consider a system of particles at rest [3] in the before zone. As we have seen, their geodesics in BJR coordinates are given by

$$\begin{cases} \hat{\mathbf{x}} = \hat{\mathbf{x}}_0 \\ \hat{v} = W(\hat{u} - \hat{u}_0 + \hat{v}_0), \end{cases} \tag{4.71}$$

with $\hat{\mathbf{x}}_0$ and $\hat{v}_0$ the initial values. For the flat metric (4.35), with the matrix $c_0 \neq 0$,

---
[3] They can represent the detectors, i.e., when one would to study the gravitational memory effect.

the matrix (4.38) is

$$H(u) = -a_0^{-1/2} c_0^{-1} \left[ (1 + (u - u_0)c_0)^{-1} - 1 \right] a_0^{-1/2} . \tag{4.72}$$

Given the (4.72), one can calculate the quantities $p_l$, $p_b$ and $d$, and finds

$$\begin{cases} p_l = -a_0^{1/2} c_0 \hat{\mathbf{x}}_0 \\ p_b = a_0^{-1/2} \hat{\mathbf{x}}_0 \\ d = \hat{v}_0 - W\hat{u}_0 + \frac{1}{2}\hat{\mathbf{x}}_0 \cdot c_0 \hat{\mathbf{x}}_0 . \end{cases} \tag{4.73}$$

Starting from (4.73), it is possible to express the geodesics (4.55), (4.59) and $u$ in the original BJR coordinates, as

$$\begin{cases} \mathbf{x}(u) = \left[ -H(u)a_0^{1/2} c_0 + a_0^{-1/2} \right] \hat{\mathbf{x}} \\ u = \hat{x} \\ v(u) = \hat{v} + \frac{1}{2}\hat{\mathbf{x}} \cdot \left[ c_0 - c_0 a_0^{1/2} H(u) a_0^{1/2} c_0 \right] \hat{\mathbf{x}} . \end{cases} \tag{4.74}$$

Using the matrix $H$ given in (4.38), we can extend the geodesics (4.74) in the sandwich. The metric in the new BJR coordinates system (4.74) is

$$g = d\mathbf{x} \cdot a(u)d\mathbf{x} + 2dudv = d\hat{\mathbf{x}} \cdot \hat{a}(u)d\hat{\mathbf{x}} + 2d\hat{u}d\hat{v} , \tag{4.75}$$

with

$$\hat{a}(u) = \left( a_0^{-1/2} - c_0 a_0^{1/2} H(u) \right) a(u) \left( a_0^{-1/2} - H(u) a_0^{1/2} c_0 \right) . \tag{4.76}$$

In conclusion, we have seen how the passages from the BJR coordinates to Brinkmann (B) coordinates, and *viceversa*, and the use of the quantities which emerge in the study of the isometries of the plane gravitational waves, allow to highlight some interesting results. It is suitable to underline the fact that the descriptions in B coordinates and in BJR coordinates are consistent: from the numerical calculations one verifies that pushing forward to B coordinates, a solution in BJR coordinates yields a trajectory which coincides with the one calculated independently in B coordinates, in the region of the spacetime where the BJR metric have no singularities.

### 4.2.2 BJR pp-wave and the Carroll group: an example

The aim of this subsection is to patch together the last arguments, namely the action of the Carroll group on a geodesic and the evolution of the profile of a plane gravitational wave in BJR coordinates.
Let us consider, i.e., the matrix $P$ in (4.27) of the form

$$P(u) = \chi(u) diag(e^{k(u)}, e^{-k(u)}). \tag{4.77}$$

The metric (4.30) becomes

$$g = \chi^2 \left[ e^{2k}(dx^1)^2 + e^{-2k}(dx^2)^2 \right] + 2dudv \tag{4.78}$$

and is Ricci-flat if it satisfies the (4.67).
In this case, the pp-wave profile is

$$K_{ij}(U)X^i X^j = \frac{1}{2} A(U)[(X^1)^2 - (X^2)^2], \quad A = \frac{2}{\chi^2} \frac{d\alpha}{du} \left( \chi^2 \frac{d\alpha}{du} \right). \tag{4.79}$$

In classical mechanics the (4.79) describes two uncoupled harmonic oscillators with opposite frequency-squares, hence one attractive and the other repulsive.

The metric (4.78) is manifestly invariant under the transverse space translations and the advanced time translation, $\mathbf{x} \to \mathbf{x} + \mathbf{l}$ and $v \to v + f$; while, the retarded time is fixed.
Since the the spatial metric is not diagonal unless $k(u) = 0$, the orthogonal group $O(2)$ is broken. Thus, the Carroll group acts as we have seen previously. In this case, the matrix $H$ has the following expression

$$H(u) = \int_{u_0}^{u} P(\varrho)^{-2} d\varrho,. \tag{4.80}$$

With the choice $k(u) = u$, a Ricci-flat metric which is regular in the neighborhood of $u = 0$ is given, for example, by $\chi(u) = -\cos u$, and then the profile of the wave is

$$\frac{1}{2} K_{ij}(U)X^i X^j = \tan U[(X^2)^2 - (X^1)]^2. \tag{4.81}$$

It describes the saddle-like surface depending on the retarded time variable $u$.

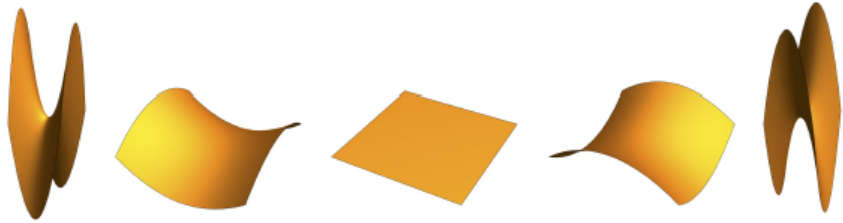

Figure 4.3: Wave profile for $k(u) = u$ and $u = \pi/2 + 0.1$, $u = -\pi/4$, $u = 0$, $u = \pi/4$, $u = \pi/2 - 0.1$. The $\pm 0.1$ pieces is due to the fact that $u = \pm\pi/2$ are two singular values for the metric, in the sense given in the last subsection [14].

In figure (4.3) is shown the wave profiles corresponding to some $u$ passing from negative to positive values.

The components of the matrix $H = diag(H_{11}, H_{22})$, which determines both the Carroll group action and the evolution of the geodesics, are

$$H_{11}(u) = \int_0^u \cos^{-2}\varrho\, e^{-2\varrho} d\varrho, \qquad H_{22} = \int_0^u \cos^{-2}\varrho\, e^{2\varrho} d\varrho. \qquad (4.82)$$

In figure (4.4), it is possible to see, for $u > 0$, that $H_{22}$ increases rapidly, while $H_{11}$ is damped. For $u < 0$ the behaviour is opposite, consistent with the change of the profile of figure (4.3). Furthermore, the change of the profile is consistent with the evolution of the geodesics, as we can see in figure (4.5a); the repulsive and attractive directions are interchanged when $u$ changes sign. In figure 4.5b) is shown the action of the Carroll group on the original curling trajectory which becomes straightened, under the transformations $p_l \to 0$, $p_b \to 0$ and $d \to 0$.

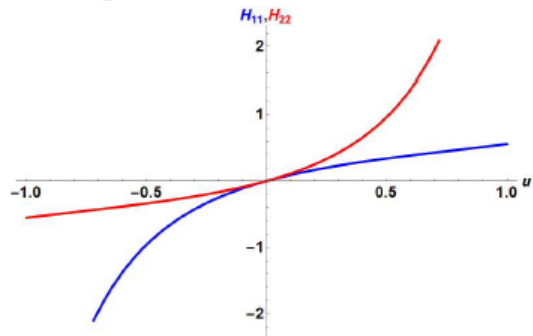

Figure 4.4: The evolution of $H_{11}$ and $H_{22}$ with $u$ [14].

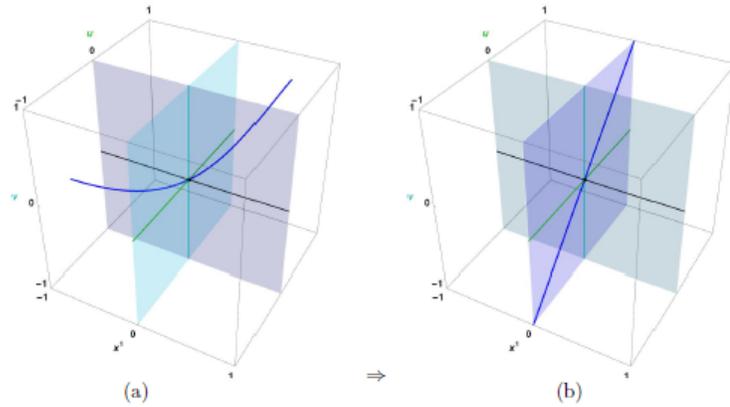

Figure 4.5: (a) original geodesic motion. (b) geodesic motion following the action of the Carroll group, with the choice $p_b = 0$ and $p_l = (0, 1)$ [14].

**Remarks**   We have omitted the discussion about some interesting examples that can highlight important physical results. For example, analyzing the geodesics in Brinkmann coordinates and, then, takes as a toy example a Gaussian gravitational burst, one can compute the displacement in the relative euclidean distance of two test particles, initially at rest [53][4]. Furthermore, one measures not only the displacement in the relative position but also a jump in the relative velocity [53] [5].

---

[4]The so-called gravitational memory effect [6]
[5]The so-called non-linear memory gravitational effect [55].

# Chapter 5

# Conclusions

This thesis had the purpose to analyze the hot topic of the asymptotic symmetries of an asymptotically flat spacetime. The main interest was devoted to the comparison between the BMS and other groups, in particular the Carroll group. It was shown that the conformal extension of the Carroll group is isomorphic to the BMS group. This fact could be useful for a deeper understanding of the BMS group, in particular about the problem of its representations, which was no matter of discussion of this thesis. The attention in the last chapter was devoted to the identification of $\mathscr{I}$ with a Carroll structure and to the study of the isometries of plane gravitational waves. The latter application allows the study of the motion of freely falling particles and the calculation of the linear memory effect in BJR coordinates. Then, one can think to replace in the infrared triangle the conformal Carroll group instead of the BMS group, but the precise link between the Carroll group and the quantistic counterpart of the triangle, the soft theorems, is still unknown. Further investigations on the Carroll group could produce other interesting results.

# Appendix A

# Differential geometry

## A.1 Differentiable manifold

**Definition 1.** (Chart) Given a space $M$, a *chart* on $M$ is a $1-1$ map from an open subset $U \subset M$ to an open subset $\phi(U) \subset \mathbb{R}^n$, i.e., a map $\phi : M \to \mathbb{R}^n$. A chart is often called a coordinate system.

In order to give the definition of a differentiable[1] manifold, we clarify the concepts of related-chart and atlas. Two charts $\phi_1, \phi_2$ are $C^\infty$-related if both the map $\phi_2 \circ \phi_1^{-1} : \phi_1(U_1 \cap U_2) \to \phi_2(U_1 \cap U_2)$ and its inverse are $C^\infty$. A collection of $C^\infty$-related charts such that every point on $M$ lies in the domain of at least one chart forms an *atlas*.

**Definition 2.** (Differentiable manifold) If $M$ is a space and $A$ its maximal atlas, the set $(M, A)$ is a $(C^\infty-)$ *differentiable manifold*.

## A.2 Tangent vectors and tangent spaces

We start by defining a curve within a manifold.

**Definition 3.** (Curve) A $C^\infty$-curve in a manifold $M$ is a map $\lambda$ of the open interval $I = (a, b) \in \mathbb{R} \to M$ such that for any chart $\phi$, $\phi \circ \lambda : I \to \mathbb{R}^n$ is a $C^\infty$ map.

The definition implies that the curve is parametrized, i.e., by a parameter $t$, then the curve is $\lambda(t)$ with $t \in (a, b)$.

---

[1] Most concepts in physics involve the notion of differentiability

**Definition 4.** (Tangent vector) The *tangent vector* $\dot{\lambda}_p = (d\lambda/dt)_p$ to a curve $\lambda(t)$ at a point $p$ on it is the map from the set of real functions $f$ defined in a neighbourhood of $p$ to $R$, defined by

$$\lambda_p : f \to \left[\frac{d}{dt}(f \circ \lambda)\right]_p = (f \circ \lambda)\dot{}_p = \dot{\lambda}_p(f).$$

The components of $\dot{\lambda}_p$ with respect to the chart are

$$(x^i \circ \lambda)\dot{}_p = \left[\frac{d}{dt}x^i(\lambda(t))\right]_p.$$

The set of the tangent vectors at $p$ constitutes the *tangent space* $T_p(M)$ at $p$.

**Definition 5.** (Dual space) The *dual space* $T_p^*(M)$ of $T_p(M)$ is the vector space of linear maps $\lambda : T_p(M) \to \mathbb{R}$.

The dual space $T_p^*(M)$ has the same dimensions of $T_p(M)$ and its elements are called 1-*forms* or *covectors*. Let $f$ be any function on $M$. For each $X_p \in T_p(M)$, $X_p f$ is a scalar. Thus $f$ defines a map $df : T_p(M) \to \mathbb{R}$ via

$$df(X_p) = X_p f,$$

where $df$ is the differential of gradient of $f$.

Let us consider the union of tangent spaces $T_p(M)$ at all points $p \in M$

$$T(M) = \cup_{p \in M} T_p(M).$$

**Definition 6.** (Vector field) A *vector field* on $M$ is a map $X : M \to T(M)$ such that $X(p) = X_p$ is a vector in $T_p(M)$.

Analogous definitions hold for covector and more general tensor fields.

## A.3 Maps and Manifold

**Definition 7.** (Map) A map $h : M \to N$ is $C^\infty$ if for every $C^\infty$ function $f : N \to \mathbb{R}$ the function $fh : M \to \mathbb{R}$ is also $C^\infty$.

**Definition 8.** (Push-forward) Given a $C^\infty$ map $h : M \to N$, the *push forward* is a map $h_* : T_p(M) \to T_{h(p)}(N)$ which maps the tangent vector of a curve $\gamma$ at $p \in M$ to the tangent vector of the curve $h(\gamma)$ at $h(p) \in N$.

For each $C^\infty$ function $f : N \to R$

$$(h_* X_p)f = X_p(f \circ h)$$

and $h_*$ is a linear map.

**Definition 9.** (Pull-back) The maps $h$ and $h_*$, previously defined, induce a linear map, called *pull-back*, $h^* : T^*_{h(p)}(N)$ such that, if $\omega \in T^*_{h(p)}(N)$, $X_p \in T_p(M)$, then

$$(h^*\omega)(X_p) = \omega(h_* X_p).$$

## A.4  Integral curves and Lie derivatives

An integral curve of a vector field $X$ in $M$ is a curve in $M$ such that at each point $p$ on $\gamma$, the tangent vector is $X_p$. It is possible to demonstrate that every smooth vector field $X$ defines locally a unique integral curve $\gamma$ through each point $p$ such that $\gamma(0) = p$.

**Definition 10.** (Lie derivative) Let $p$ be a point of $M$ and let $X$ be a smooth vector field. Let $\gamma$ be the integral curve of $X$ through $p$ inducing a 1-parameter group of transformations $(h_t)$. Then if $f : M \to \mathbb{R}$ is any real function on $M$, the *Lie derivative* of $f$ with respect to $X$ is

$$(\mathcal{L}_X f)_p = \lim_{dt \to 0} \left[ \frac{f(h_{dt}(p)) - f(p)}{dt} \right].$$

For any function $f$ on $M$

$$(\mathcal{L}_X f)_p = X_p f.$$

The Lie derivative of a smooth vector field $Z$ with respect to $X$ at $p$ is given by

$$(\mathcal{L}_X Z)_p = [X, Z]_p.$$

Now, given an arbitrary vector field $V^a$, the Lie derivative w.r.t. the vector field of the metric tensor $g_{ab}$ is

$$\mathcal{L}_V g_{ab} = V^c \nabla_c g_{ab} + g_{cb} \nabla_a V^c + g_{ac} \nabla_b V^c = \nabla_a V_b + \nabla_b V_a \,.$$

## A.5 Killing vector fields

**Definition 11.** (Killing vector field) Let us $\phi_t : M \to M$ a one-parameter group of isometries, namely $\phi_t^* g_{ab} = g_{ab}$, the vector field $\xi^a$ which generates $\phi_t$ is called a *Killing vector field* .

The map $\phi_t$ is an isometry group iff $\mathcal{L}_\xi g_{ab} = 0$. Thus, the necessary and sufficient condition for $\xi^a$ to be a Killing vector is that it satisfy Killing's equation

$$\nabla_a \xi_b + \nabla_b \xi_a = 0 \,,$$

where $\nabla_a$ is the covariant derivative associated to $g_{ab}$.

## A.6 Linear connections

**Definition 12.** (Linear connection) A *linear connection* $\nabla$ on $M$ is a map sending every pair of smooth vector fields $(X, Y)$ to a vector field $\nabla_X Y$ such that

$$\nabla_X(aY + Z) = a\nabla_X Y + \nabla_X Z \,,$$

for any constant scalar a, but

$$\nabla_X(fY) = f\nabla_X Y + (Xf)Y \,,$$

when $f$ is a function, and it is linear in $X$

$$\nabla_{X+fY} Z = (\nabla_X Z + f\nabla_Y Z) \,.$$

Further, acting on functions $f$, $\nabla_X$ is defined by

$$\nabla_X f = Xf \,.$$

The operator $\nabla_X Y$ is called the covariant derivative of $Y$ with respect to $X$. Because $\nabla_X Y$ is not linear in $Y$, $\nabla$ is not a tensor.

Let $(e_a)$ be a basis for vector fields and write $\nabla_{e_a}$ as $\nabla_a$. Since $\nabla_a e_b$ is a vector there exist scalars $\Gamma^c_{ba}$ such that

$$\nabla_a e_b = \Gamma^c_{ba} e_c,$$

where $\Gamma^c_{ba}$ are the components of the connection.

**Definition 13.** If $\nabla_X Y = 0$ then $Y$ is said to be *parallely transported* with respect to $X$.

# Appendix B

# Lie groups and Lie algebras

A Lie group is a group which is also a manifold, with the property that the group operations are compatible with the smooth structure. It represents the best theory describing the continous symmetries of a mathematical structure.

**Definition 14.** (Lie group) An $r$-parameter Lie group is a group $G$ which carries the structure of an $r$-dimensional smooth manifold in such a way that, given $g$ and $h$ two elements of $G$, both the group operation

$$m : G \times G \to G, \qquad m(g,h) = g \cdot h, \qquad g, h \in G,$$

and the inversion

$$i : G \to G, \qquad i(g) = g^{-1}, \qquad g, \in G,$$

are smooth maps between manifolds.

A Lie group *homomorphism* is a smooth map $\phi : G \to H$ between two Lie groups which respects the group operations:

$$\phi(g \cdot \tilde{g}) = \phi(g) \cdot \phi(\tilde{g}), \qquad g, \tilde{g} \in G.$$

The map $\phi$ is called an *isomorphism* between $G$ and $H$ if it has a smooth inverse.

If $G$ is a Lie group, then there are some vector fields on $G$ which are invariant under the group multiplication. These invariant vector fields form a finite-dimensional vector space, called the Lie algebra of $G$, and represent the infinitesimal generator of $G$. The *Lie algebra* of a Lie group $G$ is traditionally denoted by the lowercase German letter $\mathfrak{g}$ and is the vector space of all the right-invariant vector fields on $G$. In order to define a right-invariant vector field, we have to introduce the right multiplication map for any $g \in G$

$$R_g : G \to G$$

defined by
$$R_g(h) = h \cdot g$$

is a diffeomorphism, with inverse

$$R_{g^{-1}} = (R_g)^{-1}.$$

A vector field $v$ on $G$ is called *right-invariant* if

$$dR_g(\mathbf{v}\,|_h) = \mathbf{v}\,|_{R_g(h)} = \mathbf{v}\,|_{hg}\,.$$

**Definition 15.** (Lie algebra) A *Lie algebra* is a vector space $\mathfrak{g}$ endowed with a bilinear operation, called the *Lie bracket*

$$[\cdot,\cdot] : \mathfrak{g} \times \mathfrak{g} \to \mathfrak{g},$$

which satisfies the following axioms

1. *Bilinearity* :
$$[c\mathbf{v} + c'\mathbf{v}', \mathbf{w}] = c[\mathbf{v}, \mathbf{w}] + c'[\mathbf{v}', \mathbf{w}],$$
$$[\mathbf{v}, c\mathbf{w} + c'\mathbf{w}'] = c[\mathbf{v}, \mathbf{w}] + c'[\mathbf{v}, \mathbf{w}'],$$

   with $w$ another right invariant vector field on $G$ and $c, c' \in \mathbb{R}$;

2. *Skew-Symmetry*
$$[\mathbf{v}, \mathbf{w}] = -[\mathbf{w}, \mathbf{v}];$$

2. *Jacobi Identity*
$$[\mathbf{u}, [\mathbf{v}, \mathbf{w}]] + [\mathbf{w}, [\mathbf{u}, \mathbf{v}]] + [\mathbf{v}, [\mathbf{w}, \mathbf{u}]] = 0$$

   for all $\mathbf{u}, \mathbf{v}, \mathbf{v}', \mathbf{w}, \mathbf{w}'$ in $\mathfrak{g}$.

We consider now a *local* Lie group $V \subset \mathbb{R}^r$ with multiplication $m(x, y)$.

**Proposition 1.** *Given $V \subset \mathbb{R}^r$ a local Lie group with multiplication $m(x, y)$, with $x, y \in V$, then the Lie algebra $\mathfrak{g}$ of right-invariant vector fields on V is spanned by the vector fields*

$$\mathbf{v}_k = \sum_{i=1}^{r} \xi_k^i(x) \frac{\partial}{\partial x^i}, \qquad k = 1, ..., r$$

*where*

$$\xi_k^i(x) = \frac{\partial m^i}{\partial x^k}(0, x).$$

**Central Extension**  A Lie algebra $\mathfrak{a}$ is called *abelian* if the Lie bracket of $\mathfrak{a}$ is trivial, that is $[\mathbf{v}, \mathbf{w}] = 0$ for all $\mathbf{v}, \mathbf{w} \in \mathfrak{a}$.

**Definition 16.** Let $\mathfrak{a}$ be an abelian Lie algebra over $\mathbb{K}$ and $\mathfrak{g}$ a Lie algebra over $\mathfrak{K}$. An exactsequence of Lie algebra homomorphisms

$$0 \to \mathfrak{a} \to \mathfrak{h} \to \mathfrak{g} \to 0$$

is called a *central extension* of $\mathfrak{g}$ by $\mathfrak{a}$, if $[\mathfrak{a}, \mathfrak{h}] = 0$, that is $[\mathbf{x}, \mathbf{y}] = 0$ for all $X \in \mathfrak{a}$ and $Y \in \mathfrak{h}$. Here we identify $\mathfrak{a}$ with the corresponding subalgebra of $\mathfrak{h}$. For such a central extension the abelian Lie algebra $\mathfrak{a}$ is realized as an ideal in $\mathfrak{h}$ and the homomorphism $\pi : \mathfrak{h} \to \mathfrak{g}$ serves to identify $\mathfrak{g}$ with $\mathfrak{h}/\mathfrak{a}$.